\documentclass[aip,pof,reprint,amsmath,amssymb,onecolumn]{revtex4-1}

\usepackage{graphicx}
\usepackage{dcolumn}
\usepackage{bm}
\usepackage{color}

\begin{document}

\title[Beating the jetting regime]{Beating the jetting regime}

\author{Alban Sauret}
\affiliation{Institut de Recherche sur les Ph\'enom\`enes Hors \'Equilibre,
CNRS/Universit\'es Aix-Marseille,\break
49, rue F. Joliot-Curie, BP 146, F-13384 Marseille cedex 13, France \\}
\affiliation{Department of Mechanical Engineering, University of Hong Kong, Pokfulam Road}
\author{Ho Cheung Shum}
\affiliation{Department of Mechanical Engineering, University of Hong Kong, Pokfulam Road}

\date{2012}

\begin{abstract}
We study numerically the dynamics of jets and drops in a microcapillary co-flow device. The co-flow stream encounters different flow regimes, including dripping, jetting. Using a level-set/finite element axysimmetric numerical simulation, we study the dynamics of breakup of a jet subject to flow rate perturbations. A scaling law for the width of the unperturbed jet is presented and compared to existing experimental results as well as numerical measurements. Then, we show that the introduction of a sinusoidal perturbation of the inner fluid flow rate can facilitate breakup of the fluid in regimes where a jet is usually observed. Moreover, the flow rate perturbation leads to a good control over the size and the frequency of the resulting droplets. Using theoretical modelling, we provide a criterion to determine the optimal frequency to break up the jet. We also derive scaling laws to determine the volume of inner fluid encapsulated in the emulsion droplets as a function of the frequency and to estimate the distance for the jet to break up as a function of the amplitude of perturbation. These scaling laws are in good agreement with results of numerical simulations. Our work suggests a novel approach and offers guiding principles to break up liquid jets in cases where dripping is difficult to achieve.
\end{abstract}

\maketitle

\section{Introduction}

The dynamics of a jet of one dispersed phase injected in a co-flowing continuous phase constitutes an interesting problem in fluid mechanics and is of fundamental importance in industrial applications that involve emulsions and bubbles. Indeed, a particular use of this system is to generate emulsions, which consist of droplets of one inner phase dispersed in a second immiscible fluid, called outer or continuous phase. The resulting multiphase flow and droplets are widely used in the industries of food, pharmaceutics, cosmetics and bio-medicine. Recent advances in microfluidics have enabled generation of emulsions with more sophisticated geometries. For instance, multiple emulsion can now be formed with a high degree of control over the size uniformity and a perfect encapsulation efficiency \cite{utada2005}.

To generate monodisperse emulsion, a proven microcapillary device consists of two capillaries aligned coaxially. Improvements in the generation of simple emulsion using this method cannot be done without a good understanding of fluid dynamics at a microscopic scale. We can separate the dynamic of co-flow streams between dripping and jetting. In the dripping regime, drops are directly formed at the tip of the inner capillary and the resulting emulsion drops are monodisperse in size. In the jetting regime, a jet is formed at the tip of the inner capillary; the jet often breaks up into droplets at certain distance from the inner capillary tip. The size of the resulting droplets is usually non-uniform and different from the size of the inner tip \cite{utada2007,utada2008}. The mechanism of the dripping and jetting regimes are quite well understood; in particular the transition between these two regimes is well described by a linear stability analysis of the Rayleigh-Plateau instability in a co-flow stream \cite{guillot2007,guillot2008b}. The most unstable wavelength predicted by this analysis is in agreement with experimental results and in particular the transition from dripping to jetting is well-predicted. A number of groups attempt to produce droplets of a controlled size in the jetting regime. A useful tool to study this flow dynamics is numerical simulation, which enables simulation of the different regimes encountered in co-flow device with arbitrary geometries \cite{hua2007}. Moreover, different physical values can be extracted from the numerical simulations, such as the pressure at a given point of the flow or the velocity field, which are often difficult to measure in the experiments.

A jet of a dispersed fluid can break up into droplets due to a Rayleigh-Plateau instability \cite{rayleigh1879,plateau1849,eggers2008}. The cylindrical jet is rendered unstable by capillarity effects; this instability is influenced by parameters such as viscosity of the outer fluid or inertia. A dynamic description of the system suggests a dependence of the destabilization of the jet on the frequency. Depending on the nature of the instability, convective or absolute, a dripping or a jetting regime prevails \cite{guillot2007}. The dynamics of co-flow jet has been studied experimentally \cite{utada2007,utada2008} and the transition between dripping and jetting has been shown to depend on two dimensionless numbers. The Weber number of the inner phase is defined as:
\begin{equation}
W\!e_{in}=\frac{\rho_{in}\,r_{jet}\,{u_{in}}^2}{\gamma}
\end{equation}
\noindent where $\rho_{in}$ and $u_{in}$ denote, respectively, the density and the average velocity of the dispersed phase, $r_{jet}$ denotes the radius of the jet and $\gamma$ is the interfacial tension between the two phases. $W\!e_{in}$ represents the ratio between the inertial and interfacial tension effects. The dynamic of the flow also depends on the capillary number of the continuous phase
\begin{equation}
C\!a_{out} = \frac{\eta_{out}\,u_{out}}{\gamma},
\end{equation}
\noindent where $\eta_{out}$ and $u_{out}$ are, respectively, the dynamic viscosity and the average velocity of the continuous phase. It represents the balance of viscous and interfacial tension effects. Experimentally, it has been observed that for $W\!e_{in}<2$ and $C\!a_{out}<0.2$, dripping occurs \cite{constant14}. However, numerical results suggest that the transition also depends on the viscosity ratio of the two phases $\eta_{in}/\eta_{out}$ and occurs at smaller dimensionless numbers, typically $W\!e_{in}<0.7$ and $C\!a_{out}<0.1$ \cite{homma2010}. 

In this work, we study numerically the formation of drops in devices with the co-flow geometry under conditions where a jet is normally observed. The typical parameters used in the systematic study of the flow rate perturbation are $W\!e_{in} \sim 3$ and $C\!a_{out} \sim 0.4$. This paper constitutes a first step towards the use of induced perturbation to modify the flow. A good understanding of the fluid dynamics governing the differents regimes of flow in a coflow device is essential to increase the efficiency and the modularity of the microfluidics technology for future industrial applications. It is also instrumental to some recent applications, such as generation of water-in-water emulsions, \cite{constant13,shum2012} which have provided a new platform for studying the generation of droplets. As the surface tension of these water-water systems is very low, typically $1000$ times lower than in oil-water emulsion, the generation of droplet is impossible by classical methods. The values of the dimensionless numbers $W\!e_{in}$ and $C\!a_{out}$ observed in these systems typically lie in the range used in this paper. At even lower values of these dimensionless numbers, the initial perturbation on the interface sometimes lead to a corrugated jet as observed in three-phases flow \cite{shum2010}. A method to break the inner jet into droplets relies on pressure perturbation of the dispersed phase, leading to a monodisperse emulsion. However, the dynamics of this mechanism is not clearly understood from a fluid mechanical point of view. Thus, it is of interest to investigate the mechanism of droplet formation and the effects of the different parameters on the resulting size of the droplets as well as the flow rates that can be used for industrial applications.

The paper is organized as follows. In section II, the governing equations and the mathematical description of the system as well as the relevant different dimensionless parameters are presented. Section III is devoted to the numerical simulation: we introduce the numerical procedure used for solving the two-phase coflow problem and we validate the method by comparison with previous results obtained in the literature. Then, in section IV, we show that the radius of the unperturbed jet can be obtained from a simple scaling law, which is confirmed by our numerical simulation. In section V, we focus on the dynamic of the perturbed jet: a scaling law for the resulting size of the droplets is derived together with a condition on the amplitude of disturbance.


\section{Governing equations and mathematical description}

\subsection{Description of the system}

We consider a microcapillary coflow geometry similar to the one used in previous experimental studies, which consists of two coaxially aligned capillary tubes \cite{utada2007,utada2008,guillot2007}. The inner capillary tube is cylindrical with a tip tapered to an inner diameter of typical size $d_{tip}=20\,\mu$m and outer diameter $d_{out}=30\,\mu$m. The inner tube is placed in a surrounding square capillary with an inner diameter equal to the outer diameter of the untapered portion of the inner capillary; this ensure a good coaxial alignment \cite{utada2007}. We use another cylindrical outer capillary of inner diameter $d_{cap}=200\,\mu$m axially aligned in the square capillary as a collecting tube. We assume all the capillaries to be axisymmetrically aligned to simplify the problem and the subsequent numerical simulation.

\begin{center}
\begin{figure}
\begin{center}\includegraphics[width=10cm]{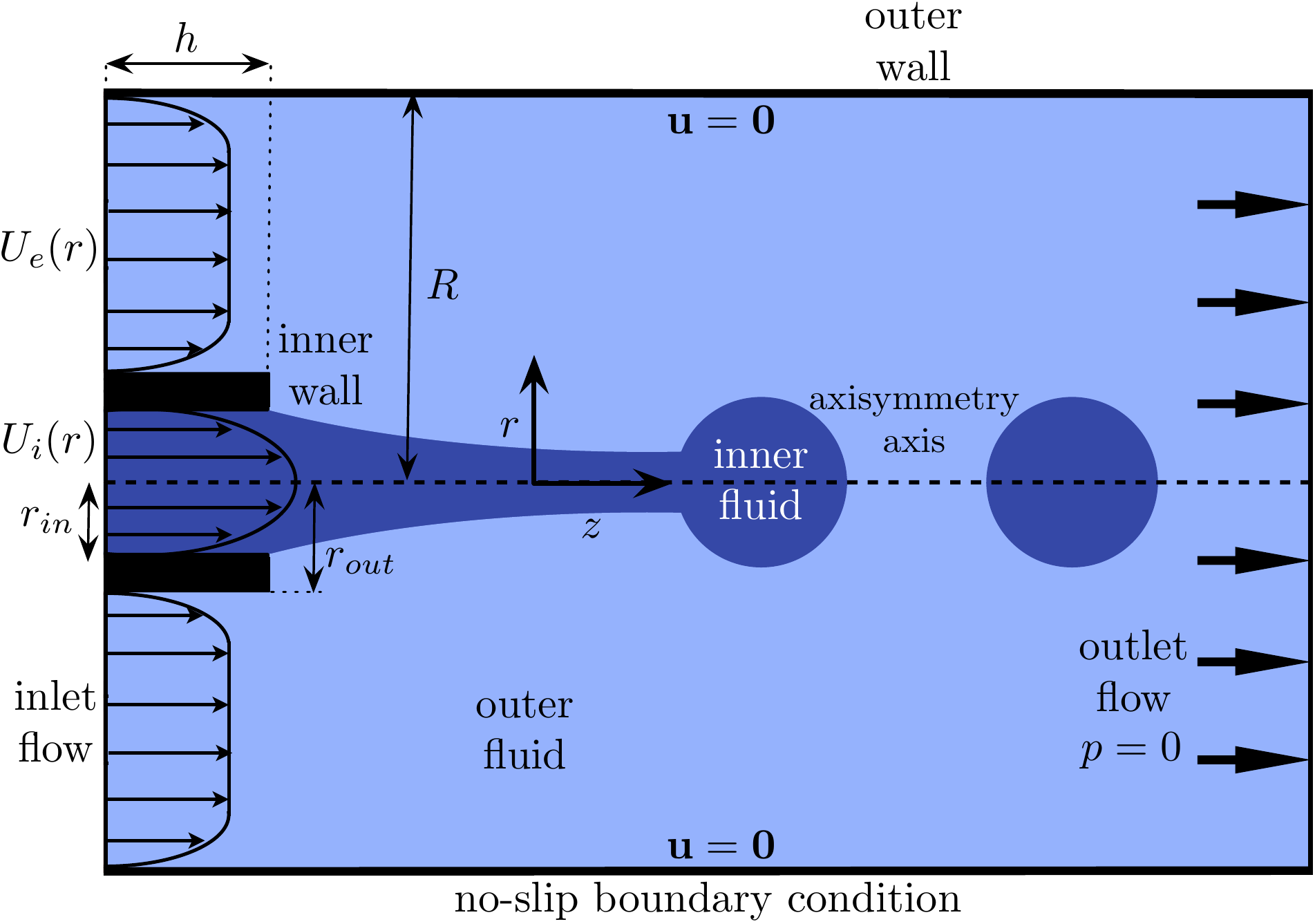}\end{center}
\caption{(Color online) Schematic of the problem. The inner fluid is injected from a capillary at the center of the domain in a coflowing outer fluid. This results in a jet that may break up into droplets downstream in the microchannel.}
\label{scheme_device}
\end{figure}
\end{center}

The experimental device leads to the numerical modelization represented schematically in figure \ref{scheme_device}. An inner fluid is injected at the center in an axisymmetric capillary tube of inner diameter $r_{in}$ and outer diameter $r_{out}$ at the constant mean velocity $U_{in}(r)$. This inner  fluid is surrounded by an outer phase injected through the axisymmetric coaxial annulus of external radius $R$ and internal radius $r_{out}$ at the mean velocity $U_{out}(r)$. Regarding the order of magnitude of the different forces, gravitational effects are negligible due to the small dimensions of the channel in this study and consequently is not taken into account in the numerical simulation. The length of the computational domain is $L=1000\,\mu$m which represents $100\,r_{in}$ and we ensure that this is sufficiently long for all the numerical computations presented in this article. Both fluids are assumed to be incompressible with a constant density of $\rho_{in}$ and $\rho_{out}$ for the inner and outer fluids repectively. Moreover, in this study, all the fluids are newtonian and have a constant viscosity $\eta_{in}$ and $\eta_{out}$ respectively. Finally, $\gamma$ is the interfacial tension between the two phases. As the two fluids are not miscible, an axisymmetric interface is formed between the two phases. We solve the axisymmetric, incompressible, unsteady, fully developed Navier-Stokes equations in cylindrico-polar coordinates $(r,\theta,z)$. The velocity field in this system of coordinate is written as $(u,v,w)$ where $u$, $v$, $w$ denote, respectively, the radial, azimuthal and axial velocity. Moreover, due to the symmetry of the system, we have $v=0$ and thus we will use only the couple of coordinates $(u,w)$.

\subsection{Governing equations}

Both inner and outer fluids are considered as incompressible. The Navier-Stokes equation and the continuity equation with variable density and viscosity are expressed as follows:
\begin{eqnarray}\label{eq:NS}
\rho\left[\frac{\partial \boldsymbol{u}}{\partial t}+(\boldsymbol{u}\cdot\boldsymbol{\nabla})\boldsymbol{u}\right] & = & -\boldsymbol{\nabla}p+\boldsymbol{\nabla}\cdot\left[\eta\left(\boldsymbol{\nabla}\boldsymbol{u}+\boldsymbol{\nabla}\boldsymbol{u}^T\right)\right] \nonumber \\
&& \qquad +\boldsymbol{\nabla}\cdot\left[\gamma\left(\boldsymbol{I}-(\boldsymbol{n}\boldsymbol{n}^T)\right)\,\boldsymbol{\delta}\right] \\
\boldsymbol{\nabla}\cdot\boldsymbol{u} & = & 0
\end{eqnarray}
where $\boldsymbol{u}=(u,v,w)$ is the velocity of the fluid, $p$ the pressure, $\rho$ the density, $\eta$ the dynamic viscosity, $\gamma$ the surface tension, $\boldsymbol{I}$ the identity matric, $\boldsymbol{\delta}$ is a Dirac delta function that is zero everywhere except at the fluid-fluid interface and $\boldsymbol{n}$ is the unit normal vector to the interface defined by
\begin{equation}
\boldsymbol{n}=\frac{\boldsymbol{\nabla}\phi}{|\boldsymbol{\nabla}\phi|}
\end{equation}
\noindent where $\phi$ is the level-set function and will be defined more precisly in the next section. We consider a two-phase flow system, and the two fluids are incompressible. It follows that the normal velocity at the interface is continuous and is expressed as
\begin{equation}
\boldsymbol{u}_{in}\cdot \boldsymbol{n}=\boldsymbol{u}_{out}\cdot \boldsymbol{n}
\end{equation}
where the subscript $in$ and $out$ denote, respectively, the inner and outer fluids. Then, the boundary conditions for the stress tensor at the interface writes, for the normal and tangential stress respectively,
\begin{eqnarray}
\left[-p+\eta\left(\boldsymbol{\nabla}\boldsymbol{u}+\boldsymbol{\nabla}\boldsymbol{u}^T\right)\right]\cdot\boldsymbol{n}& = & \gamma\,\mathcal{C}\,\boldsymbol{n} \\
\left[\eta\left(\boldsymbol{\nabla}\boldsymbol{u}+\boldsymbol{\nabla}\boldsymbol{u}^T\right)\right]\cdot\boldsymbol{t} & = & \boldsymbol{0}
\end{eqnarray}
where $\mathcal{C}$ represents the curvature of the interface and $\boldsymbol{t}$ the unit tangent vector at the interface. 

On the solid boundaries of the capillary walls, denoted as $\Sigma$, the no-slip condition implies
\begin{equation}
\boldsymbol{u}_\Sigma=\boldsymbol{0}
\end{equation}


\subsection{Inlet and outlet flow}

Using the cylindrico-polar coordinate $(r,\theta,z)$ where the velocity field is expressed as ($u$,$v$,$w$), the inner fluid is injected through the inner capillary at the mean velocity $w_{m,in}$, which is related to the inner flow rate $Q_{in}$ by the relation
\begin{equation}
w_{m,in}  =  \frac{Q_{in}}{\pi\,{r_{in}}^2}
\end{equation}
We implement in the numerical model the mean velocity $w_{m,in}$ as the inner inlet laminar flow. Then, the inner fluid at the outlet of the capillary adopts a Poiseuille profile, which can be expressed as
\begin{equation} \label{w_in}
\left\{
    \begin{split}
    u_{in}(r) & =0 \\ 
    w_{in}(r)& =  \frac{2\,Q_{in}}{\pi\,{r_{in}}^2}\left[1-\frac{r^2}{{r_{in}}^2}\right]
    \end{split}
  \right.
    \quad \text{for} \quad r\in[0;r_{in}]
\end{equation}
The inner fluid velocity profile satisfies the no-slip condition at $r=r_{in}$. The inner flow rate is
\begin{equation}\label{debit_out}
Q_{in}=\int_{0}^{r_{in}}2\pi\,r\,w_{in}(r)\,\text{d}r
\end{equation}

The average velocity in the outer annular channel $w_{m,out}$ is given by
\begin{equation}
w_{m,out}  =  \frac{Q_{out}}{\pi\,(R^2-{r_{out}}^2)}
\end{equation}
For the outer inlet laminar flow, we implement in the numerical model the mean velocity $w_{m,out}$. The outer fluid in the outer annular channel has a more complicated profile given by
\begin{eqnarray}\label{w_out}
& &\left\{
    \begin{split}
    u_{out}(r) & =0 \\ 
   \displaystyle w_{out}(r)& =  \frac{2\,Q_{out}}{\pi\,(R^2-{r_{out}}^2)}\,\frac{  \displaystyle 1-\frac{r^2}{R^2}-\left[\frac{1-{{r_{out}}^2}/{R^2}}{\ln\left({R}/{r}\right)}\right]\ln\left(\frac{R}{r_{out}}\right)}{  \displaystyle1+\frac{{r_{out}}^2}{R^2}-\frac{1-{{r_{out}}^2}/{R^2}}{\ln({R}/{r_{out}})}}
    \end{split}
  \right.  \nonumber\\
& &  \qquad\qquad\qquad\qquad\qquad\qquad\qquad \text{for} \quad r\in[r_{out};R]
\end{eqnarray}
which satisfies the no-slip boundary conditions at $r=r_{out}$ and $r=R$. The outer external flow rate is directly obtained by the relation:
\begin{equation}\label{debit_out}
Q_{out}=\int_{r_{out}}^{R}2\pi\,r\,w_{out}(r)\,\text{d}r
\end{equation}

\begin{center}
\begin{figure}
\begin{center}\includegraphics[width=7cm]{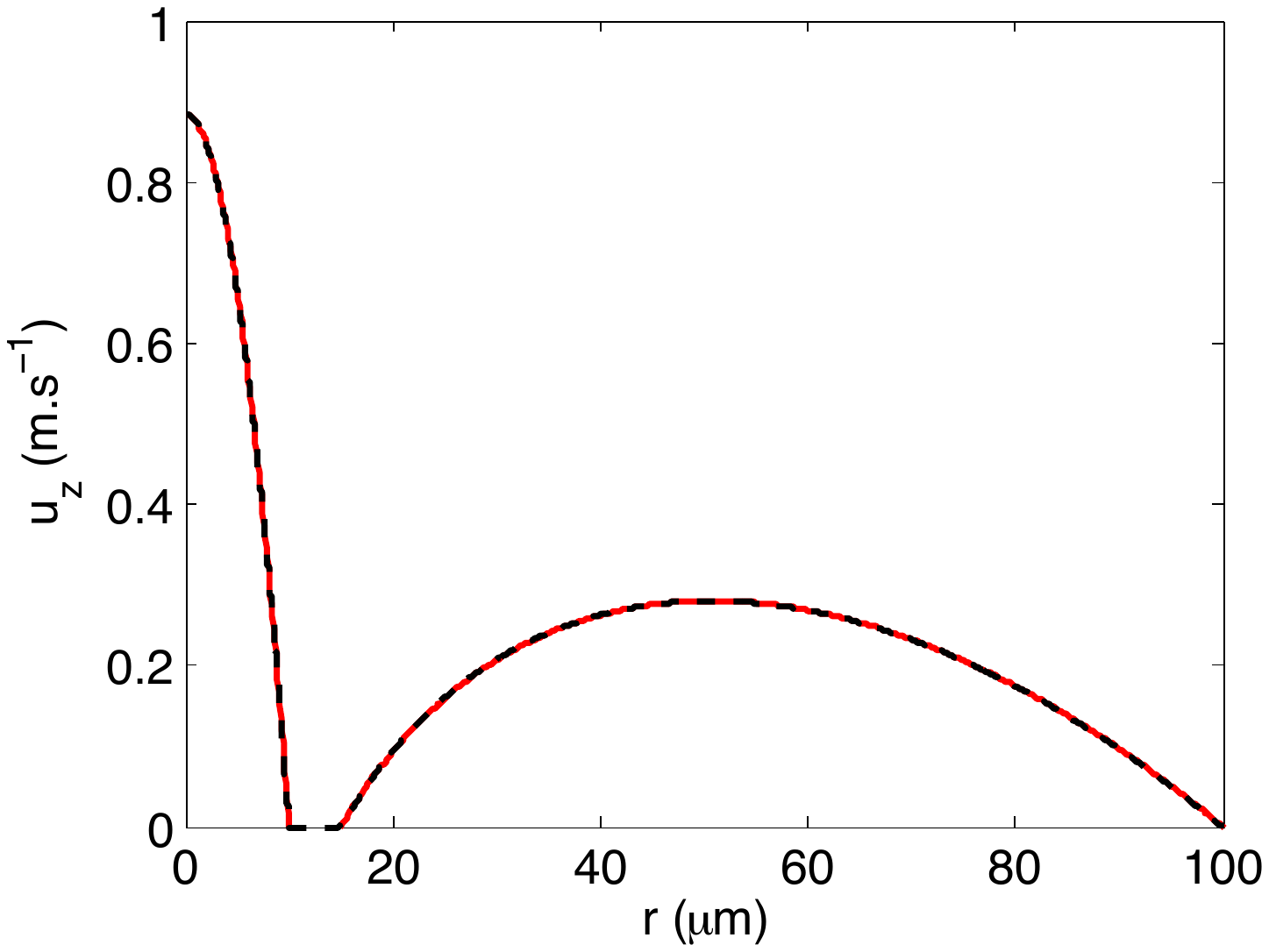}\end{center}
\caption{Comparison of theoretical axial velocity profile given by the relations (\ref{w_in}) and (\ref{w_out}) (red continuous line) and the numerical profile (black dotted line) for an inner fluid and outer fluid flow rates of $Q_{in}=500\,\mu\text{L}\cdot\text{h}^{-1}$ and $Q_{out}=20\,000\,\mu\text{L}\cdot\text{h}^{-1}$, respectively. The axial velocity is taken directly at the exit of the inner capillary.}
\label{velocity_theo_num}
\end{figure}
\end{center}

We can compare the analytical velocity profile with the numerical profile obtained at the exit of the capillary by using a mean value of the velocity for the inlet flow. The comparison is plotted in figure \ref{velocity_theo_num}, where a good agreement confirms the validation of our approach to implement the inlet flow.

Moreover, as the whole problem is axisymmetric, we solve it only in a half part of the domain presented in figure (\ref{scheme_device}) and the axis $r=0$ is taken as the axis of axisymmetry. Furthermore, an outflow boundary condition is applied at the exit of the capillary located at the right of the domain by imposing the condition of non-viscous stress together with $p=0$.

\subsection{Dimensionless parameters}

Regarding the parameters present in this problem, we can define dimensionless parameters to describe the resulting flow. These parameters are used to compare the various forces in the problem. First, the ratio of the inertia force to viscous force is characterized by the Reynolds number $Re$ defined, respectively, for the inner and outer phase by
\begin{equation}
Re_{in}  = \frac{2\,\rho_{in}\,w_{m,in}\,r_{in}}{\eta_{in}} \quad \text{and} \quad Re_{out}  =  \frac{2\,\rho_{out}\,w_{m,out}\,R}{\eta_{out}}
\end{equation}
where the prefactor of two account for the typical length of the problem, which is the diameter of the capillaries and is equal to $2\,r_{in}$ and $2\,R$ respectively. We can also define the ratio of inertia force to the surface tension effects using the Weber number
\begin{equation}
W\!e_{in}  =  \frac{2\,\rho_{in}\,{w_{m,in}}^2\,r_{in}}{\gamma} \quad \text{and} \quad W\!e_{out}  =  \frac{2\,\rho_{out}\,{w_{m,out}}^2\,R}{\gamma}
\end{equation}
Finally, the capillary number is defined by the ratio of viscous force to surface tension force:
\begin{equation}
Ca_{in}  =  \frac{\rho_{in}\,{w_{m,in}}}{\gamma} \quad \text{and} \quad  Ca_{out}  =  \frac{\rho_{out}\,{w_{m,out}}}{\gamma}
\end{equation}

We can also define the ratio of the density, $\rho_{in}/\rho_{out}$, and the ratio of dynamic viscosity, $\eta_{in}/\eta_{out}$.

\section{Numerical procedure}

\begin{center}
\begin{figure}
\begin{center}\includegraphics[width=8cm]{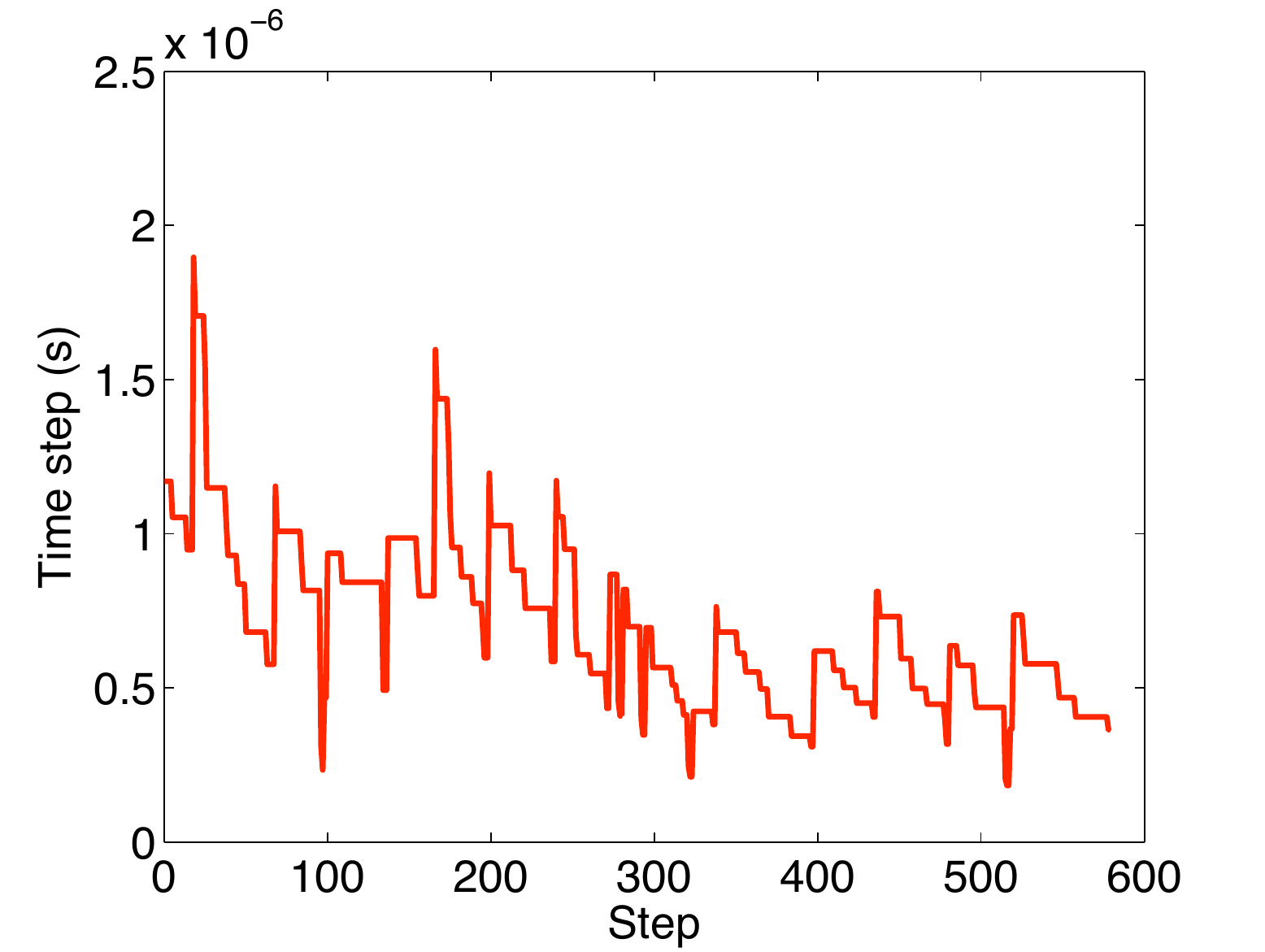}\end{center}
\caption{{Timestep for a simulation at $f=20\,000$ Hz, $W\!e_{in}=2.5$, $Ca_{out}=0.317$. The time step is always smaller than the period of the applied forcing}}
\label{step}
\end{figure}
\end{center}

We elucidate the dynamic of the two-phase flow in the microcapillary coflow system using a level set method with a finite-element numerical scheme implemetend in a commerical software, COMSOL Multiphysics$^{\copyright}$. We solve the laminar, unsteady, two-phase flow with a conservative level set method, which is used to describe the fluid interface between the two phases \cite{olsson2005,yue2004}. All simulations are perfomed using a numerical grid composed of triangular elements. All elements are of standard Lagrange $P_1-P_2$ type (i.e. linear for the pressure field and quadratic for the velocity field). The temporal solver is IDA solver (Implicit Differential-Algrebic solver), based on a backward differencing formulas \cite{hindmarsh2005}. At each time step, the system is solved with the sparse direct linear solver PARDISO\footnote{www.pardiso-project.org}. {{In this work, we are using relatively high frequency, which ranges from 600 Hz, up to 20 000 Hz. The time steps used in the simulation goes up to $T=5\times10^{-5}$ s, which is much smaller than the inverse of the applied frequency. This ensures good convergence and high precision of the numerical simulations.  As such, the numerical simulation is carried out for a sufficiently long time until a stable droplet breakup time is reached, typically around $t_{final} = 0.05$. An example of the time step obtained in the simulation at a frequency $f = 20 000$ Hz is shown in figure \ref{step}}}.

\subsection{Level-set function $\phi$}

The interface between the two fluids is defined by a level-set function, $\phi$. This function is such that in the inner phase its value is $\phi=0$ whereas in the outer phase, $\phi=1$. The function $\phi$ goes smoothly from $0$ to $1$ in a transition layer of thickness $\epsilon$. In all the simulations presented in this paper, we have used $\epsilon=h_c/2 \sim 10^{-6}$ where $h_c$ represents the typical mesh size. The interface between the inner and outer phases is defined by $\phi=0.5$ and is advected at the fluid velocity $\boldsymbol{u}$. The convection of the reinitialized level set function is described by the equation:
\begin{equation}
\frac{\partial \phi}{\partial t}+\boldsymbol{\nabla}\cdot\left(\phi\,\boldsymbol{u}\right)+\lambda\,\boldsymbol{\nabla}\cdot\left[\phi\,(1-\phi)\,\frac{\boldsymbol{\nabla}\phi}{|\boldsymbol{\nabla}\phi|}-\epsilon\,\boldsymbol{\nabla}\phi\right]=0
\end{equation}
where $\epsilon$ describes the thickness of the transition layer.  The parameter $\lambda$ determines the amount of reinitialization and its value is of the order of the maximum magnitude occurring in the velocity field.

The aim of the level set function, apart from describing the interface, is to give the density and the viscosity of the fluid at each points through the relations:
\begin{eqnarray}
\rho=\rho_{in}+(\rho_{out}-\rho_{in})\,\phi \\
\eta=\eta_{in}+(\eta_{out}-\eta_{in})\,\phi 
\end{eqnarray}

\subsection{Meshgrid}

\begin{center}
\begin{figure}
\begin{center}\includegraphics[width=10cm]{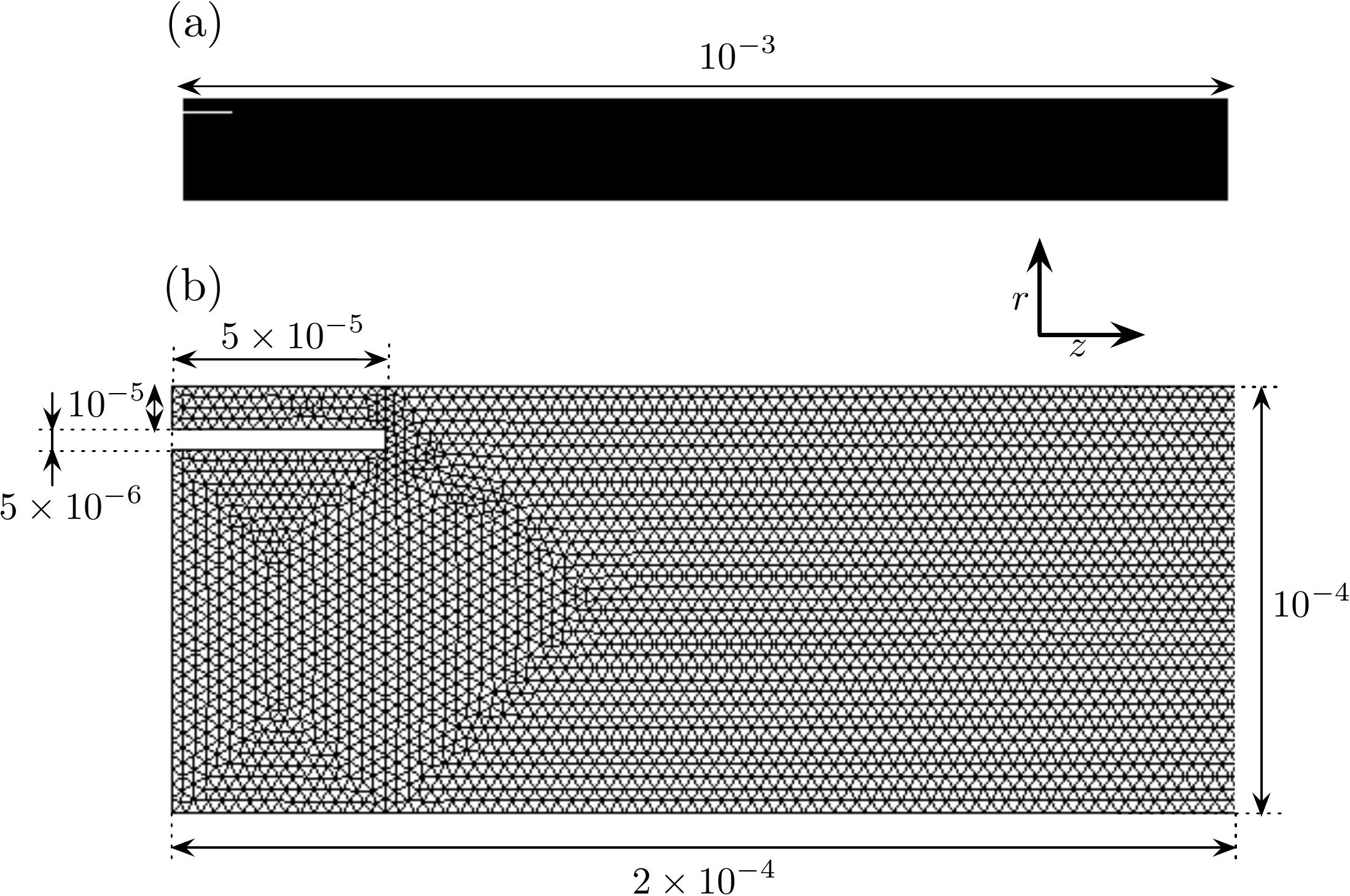}\end{center}
\caption{(a) Schematic view of the microfluidic coflow device in cylindrico-polar coordinate $(r,\theta$,z) together with the meshgrid used for the numerical simulations. The z- axis is the axis of axisymmetry. This symmetry ensures that the problem can be understood by performing the numerical calculation in only half of the system. (b) is a zoom of (a) near the left boundary where both fluids are injected; the size of the triangular mesh can be visualized in (b).}
\label{scheme_numeric}
\end{figure}
\end{center}

 In all simulation, we have used an Eulerian meshgrid which consists of triangular elements (see figure \ref{scheme_numeric}).

\begin{center}
\begin{figure}
\begin{center}\includegraphics[width=6cm]{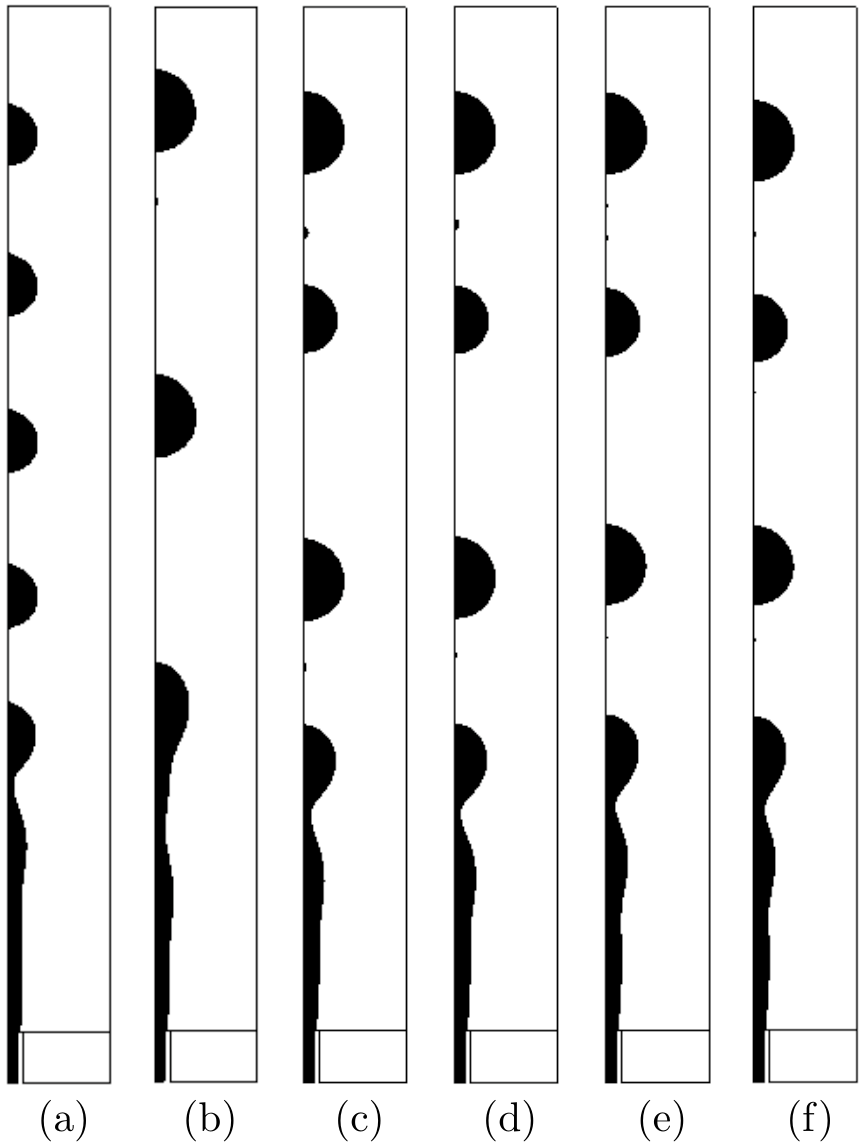}\end{center}
\caption{Comparison of the shape of the interface for $W\!e_{in}=1.5323$ and $Ca_{out}=0.0904$ and a viscosity ratio $\eta_{in}/\eta_{out}=0.1$ at the same time $t=0.082$ showing a jetting regime (a) for $17\,333$ DoF, (b) for $36\,919$ DoF, (c) for $74\,587$ DoF, (d) for $99\,977$ DoF, (e) for $143\,982$ DoF and (f) $218\,559$ DoF. The numerical results look similar beyond $74\,587$ DoF, which is sufficient for ensuring a good convergence of our numerical simulation using $143\,982$ DoF.}
\label{syst_dof}
\end{figure}
\end{center}

\begin{center}
\begin{figure}
\begin{center}\includegraphics[width=7cm]{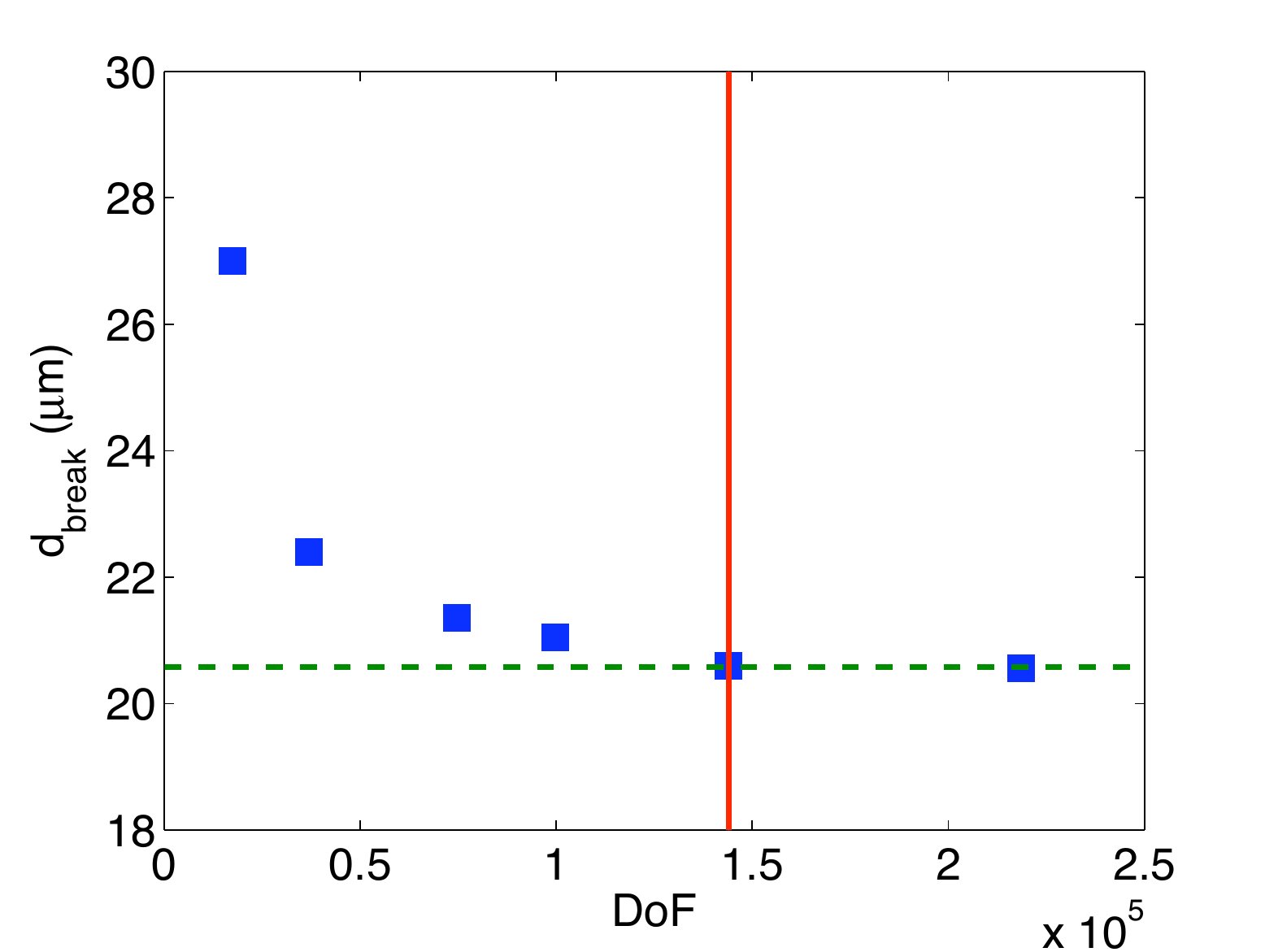}\end{center}
\caption{Distance for the jet to breakup into drops (in $\mu m$) as a function of the number of degree of freedom for $W\!e_{in}=1.5323$ and $Ca_{out}=0.0904$ and a viscosity ratio $\eta_{in}/\eta_{out}=0.1$. The distance is averaged over five breakups. The green dashed line show the value of the maximum number of DoF used, $218\,559$ DoF. The red continuous line shows the number of DoF used in the subsequent simulations ($143\,982$ DoF).}
\label{syst_dof2}
\end{figure}
\end{center}

We have performed a simulation with different number of degrees of freedom (DoF) to test the convergence of the numerical results for dimensionless parameters $W\!e_{in}=1.5323$ and $Ca_{out}=0.0904$. The results are similar for $143\,982$ DoF and higher, as shown in figure \ref{syst_dof}. This ensures the convergence of our results at $143\,982$ DoF. To compare quantitatively the influence of the number of DoF, we measure the distance for the jet to break up into drops as a function of the number of degree of freedom (see figure \ref{syst_dof2}). Using $143\,982$ DoF for the numerical simulation ensure a well-converged simulation (around $1\%$ of precision for the breakup distance). This number of DoF will be used in the following simulations; this provides the needed accuracy while keeping a reasonable computational time {{(i.e. about 4 h on a standard workstation)}}.

\subsection{Numerical validation}

\begin{center}
\begin{figure}
\begin{center}\includegraphics[width=8cm]{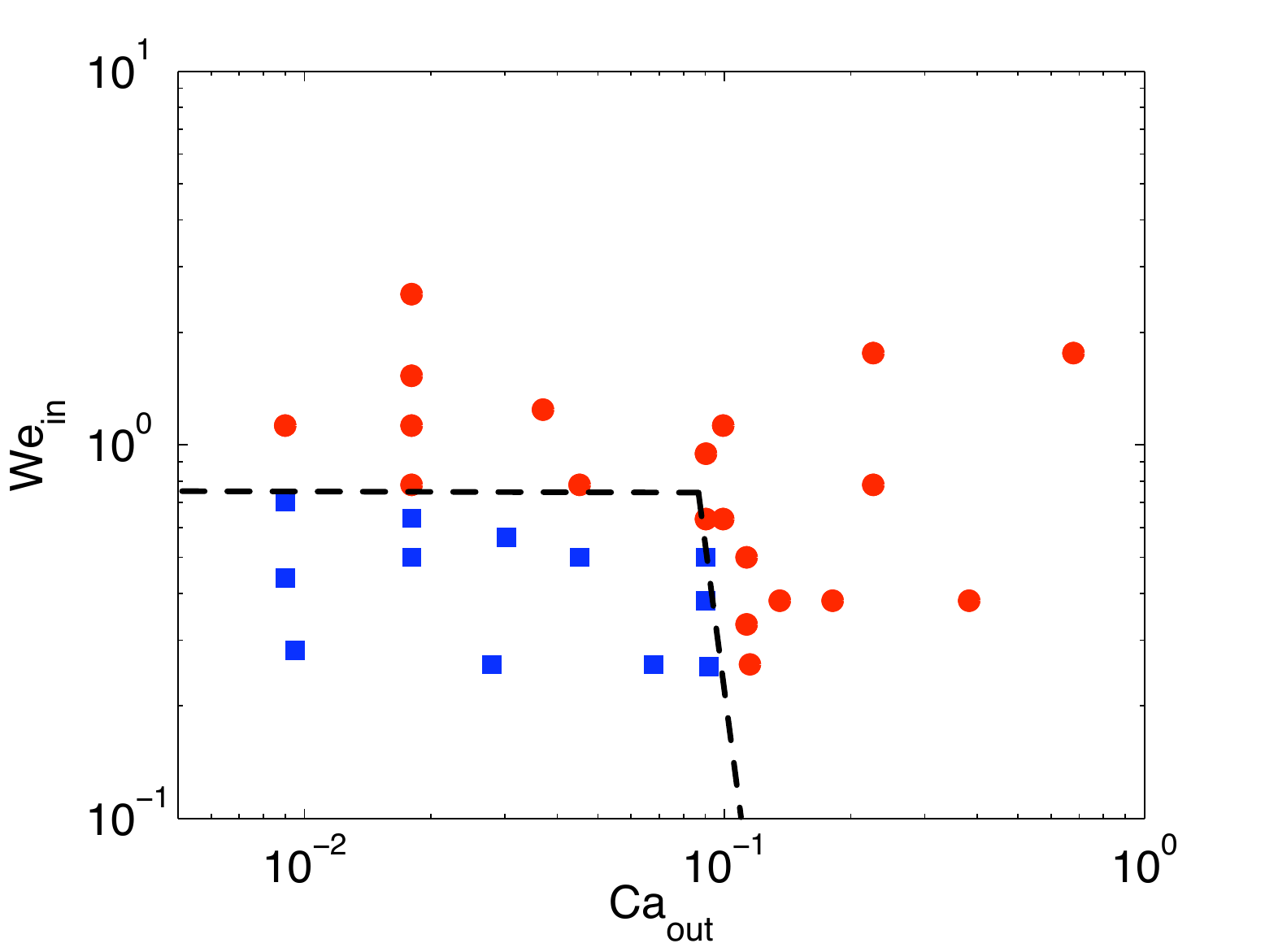}\end{center}
\caption{State diagram showing the dripping/jetting transition as a function of $Ca_{out}$ and $W\!e_{in}$ for a viscosity ratio $\eta_{in}/\eta_{out}=0.1$. Red circles indicate conditions leading to the jetting regime and blue squares indicate the dripping regime.}
\label{fig:andy_diagram}
\end{figure}
\end{center}

Even though our numerical simulation is well-converged, the results given by our numerical model should also agree with experimental data. To check this, we have performed a systematic study to determine the dripping/jetting transition in our system and compare these results to theoretical and experimental results obtained by Utada et al. for a similar system \cite{utada2007}. The results are summarized in figure \ref{fig:andy_diagram}.

We observe that the dripping/jetting transition occurs for inner Weber number, $W\!e_{in}>0.75$ and outer capillary number, $Ca_{out}>0.11$. These two values are lower than the one obtained by Utada et al. However, this discrepancy has already been observed by Homma et al \cite{homma2010} and is attributed to the difference in the viscosity ratio. Therefore, the transition obtained with our numerical model is in good agreement with previous values obtained in the literature.

\section{Size of the unperturbed jet}


We consider the axisymmetric co-flow device presented in figure \ref{scheme_device}. If no perturbation is applied and the Weber number of the inner fluid and capillary number of the outer fluid are sufficiently large, $W\!e_{in} >2$ and $Ca_{out}>0.2$ respectively, we do not observe formation of drops in a dripping regime and instead a jet is formed. Here, we evaluate the width of the undisturbed jet as a function of the different parameters including the inner and outer fluid flow rates, the viscosity of both phases, the interfacial tension and the characteristic size of the device.

The typical Reynolds number, which describes the ratio of inertial forces to viscous forces is quite low, $Re=O(1)$. Therefore, the Navier-Stokes equation is reduced in the stationary regime to
\begin{equation}\label{eq:NS}
\eta\,\boldsymbol{\nabla}^2\boldsymbol{u}=\boldsymbol{\nabla}p
\end{equation}
\noindent where $\boldsymbol{u}$ and $p$ are, respectively, the velocity and the pressure of the fluid in the considered phase. The symmetry of the problem imply that the velocity field can be expressed as
\begin{equation}
\boldsymbol{u_{out}}=w_{out}(r)\,\boldsymbol{e_z} \quad \text{and} \quad \boldsymbol{u_{in}}=w_{in}(r)\,\boldsymbol{e_z}
\end{equation}
\noindent and the pressure depends only on the axial coordinate:
\begin{equation}
p_{out}(z) \quad \text{and} \quad p_{in}(z)
\end{equation}

On the outer boundaries $r=R$, we apply the no-slip boundary condition as it remains an excellent approximation for flows at scales above micrometers \cite{stone2004}:
\begin{equation}
w_{out}(r=R)=0
\end{equation}
At the interface between the two fluids, the continuity of the velocity field is required:
\begin{equation}
w_{in}(r=r_j)=w_{out}(r=r_j)
\end{equation}
together with the continuity of the tangential shear stress
\begin{equation}
\eta_{in}\,\left[\frac{\text{d}w_{in}}{\text{d}r}\right]_{r=r_j}=\eta_{out}\,\left[\frac{\text{d}w_{out}}{\text{d}r}\right]_{r=r_j}
\end{equation}
From these boundary conditions, the equation (\ref{eq:NS}) leads to the following expressions for the velocity of the dispersed phase:
\begin{eqnarray}
w_{in}(r)=-\frac{1}{4\,\eta_{in}}\frac{\text{d}p_{in}}{\text{d}z}(r^2-{r_j}^2)+\frac{(R^2-{r_j}^2)}{4\,\eta_{out}}\frac{\text{d}p_{out}}{\text{d}z} \nonumber \\
-\frac{{r_j}^2}{2\,\eta_{out}}\left(\frac{\text{d}p_{in}}{\text{d}z}-\frac{\text{d}p_{out}}{\text{d}z}\right)\text{ln}\left(\frac{r_j}{R}\right)
\end{eqnarray}
\noindent and for that of the continuous phase:
\begin{equation}
w_{out}(r)=-\frac{1}{4\,\eta_{out}}\frac{\text{d}p_{out}}{\text{d}z}(r^2-{R}^2)-\frac{{r_j}^2}{2\,\eta_{out}}\left(\frac{\text{d}p_{in}}{\text{d}z}-\frac{\text{d}p_{out}}{\text{d}z}\right)\text{ln}\left(\frac{r}{R}\right)
\end{equation}

Therefore, the fluid flow rates of both phases, respectively, writes
\begin{eqnarray}\label{eq:flow_rate_1}
Q_{in}=2\,\pi\Bigl[\frac{{r_j}^4}{16\,\eta_{in}}\frac{\text{d}p_{in}}{\text{d}z}-\frac{{r_j}^4}{4\,\eta_{out}}\left(\frac{\text{d}p_{in}}{\text{d}z}-\frac{\text{d}p_{out}}{\text{d}z}\right)\text{ln}\left(\frac{r_j}{R}\right)\nonumber \\
+\frac{R^2-{r_j}^2}{8\,\eta_{out}}\,{r_j}^2\,\frac{\text{d}p_{out}}{\text{d}z}\Bigr]
\end{eqnarray}
\begin{eqnarray}\label{eq:flow_rate_2}
Q_{out}=\frac{\pi}{8\eta_{out}}\Bigl[\left(\frac{\text{d}p_{in}}{\text{d}z}-\frac{\text{d}p_{out}}{\text{d}z}\right)\left(4{r_j}^4\,\text{ln}\left(\frac{r_j}{R}\right)+2{r_j}^2(R^2-{r_j}^2)\right)\nonumber \\
+(R^2-{r_j}^2)^2\,\frac{\text{d}p_{out}}{\text{d}z}\Bigr]
\end{eqnarray}

The two pressure gardients can be expressed using the continuity of the normal stress at the interface
\begin{equation}
p_{in}-p_{out}=\frac{2\,\gamma}{r_j}
\end{equation}

In the unperturbed case, the radius of the jet, $r_j$, remains constant and a differentiation with respect to $z$ leads to
\begin{equation}
\frac{\text{d}p_{in}}{\text{d}z}=\frac{\text{d}p_{out}}{\text{d}z}=\frac{\text{d}p}{\text{d}z}
\end{equation}

Thus, the expressions (\ref{eq:flow_rate_1}) and (\ref{eq:flow_rate_2}) becomes
\begin{equation}
Q_{in}=2\,\pi\left[\frac{{r_j}^4}{16\,\eta_{in}}+\frac{R^2-{r_j}^2}{4\,\eta_{out}}\,{r_j}^2\right]\frac{\text{d}p}{\text{d}z}
\end{equation}
\begin{equation}
Q_{out}=\frac{\pi}{8\eta_{out}}(R^2-{r_j}^2)^2\,\frac{\text{d}p}{\text{d}z}
\end{equation}

The ratio of the inner and outer fluid flow rates writes
\begin{equation}
\frac{Q_{in}}{Q_{out}}=\frac{2\,{r_j}^2}{(R^2-{r_j}^2)}+\frac{\eta_{out}}{\eta_{in}}\frac{{r_j}^4}{(R^2-{r_j}^2)^2}
\end{equation}

Thus, we arrive at the radius of the inner jet
\begin{equation}\label{eq:scaling_jet}
r_j=R\,\left[\frac{\left(1+\displaystyle \frac{Q_{in}}{Q_{out}}\,\frac{\eta_{in}}{\eta_{out}}\right)^{1/2}-\left(1+\displaystyle \frac{Q_{in}}{Q_{out}}\right)}{\displaystyle \frac{\eta_{in}}{\eta_{out}}-2-\frac{Q_{in}}{Q_{out}}}\right]^{1/2}
\end{equation}

\begin{center}
\begin{figure}
\begin{center}\includegraphics[width=7cm]{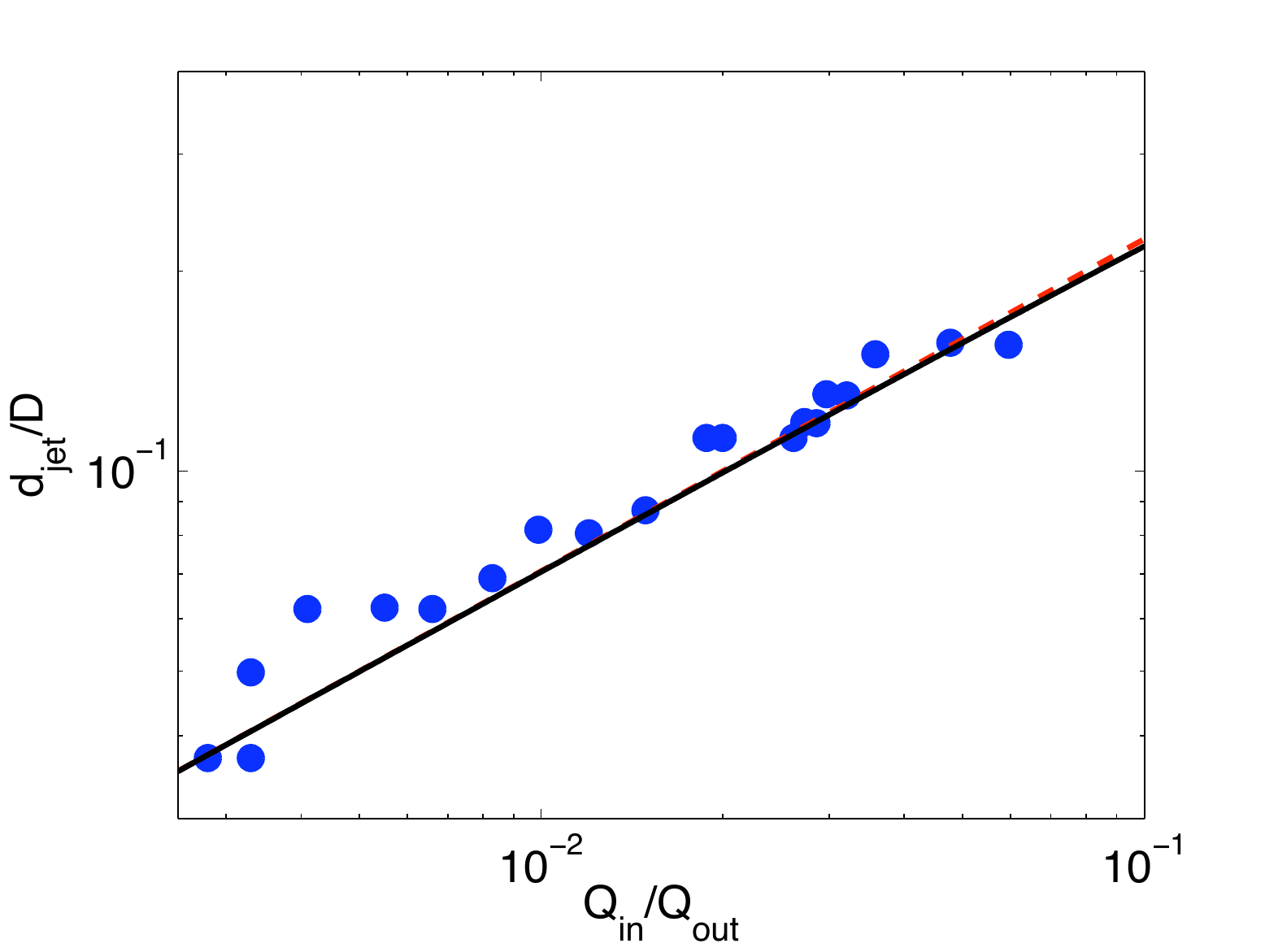}\end{center}
\caption{Diameter of the inner jet rescaled by the diameter of the outer capillary. The black line indicates the scaling law (\ref{eq:scaling_jet}) with a viscosity ratio $\eta_{in}/\eta_{out}=0.1$. The blue circles are the experimental measurements obtained by Utada et al. \cite{utada2007} together with their suggested scaling law (red dotted line).}
\label{fig:jet_radius_experiments}
\end{figure}
\end{center}

Surprisingly, we see that the scaling law suggested by Utada et al. nearly collapses on the present scaling law. Most of the experimental measurements were done at a low ratio of the inner and outer fluid flow rates, $Q_{in}/Q_{out}\ll 1$. Considering the expression (\ref{eq:scaling_jet}) in this limit lead to an asymptotic scaling law:
\begin{equation}\label{eq:scaling_jet2}
r_j=R\,\left[\frac{Q_{in}}{2\,Q_{out}}\right]^{1/2}
\end{equation}
\noindent which is exactly the scaling law derived by Utada et al. In this case, the influence of the viscosity of both phases is negligible.

\begin{center}
\begin{figure}
\begin{center}\includegraphics[width=8cm]{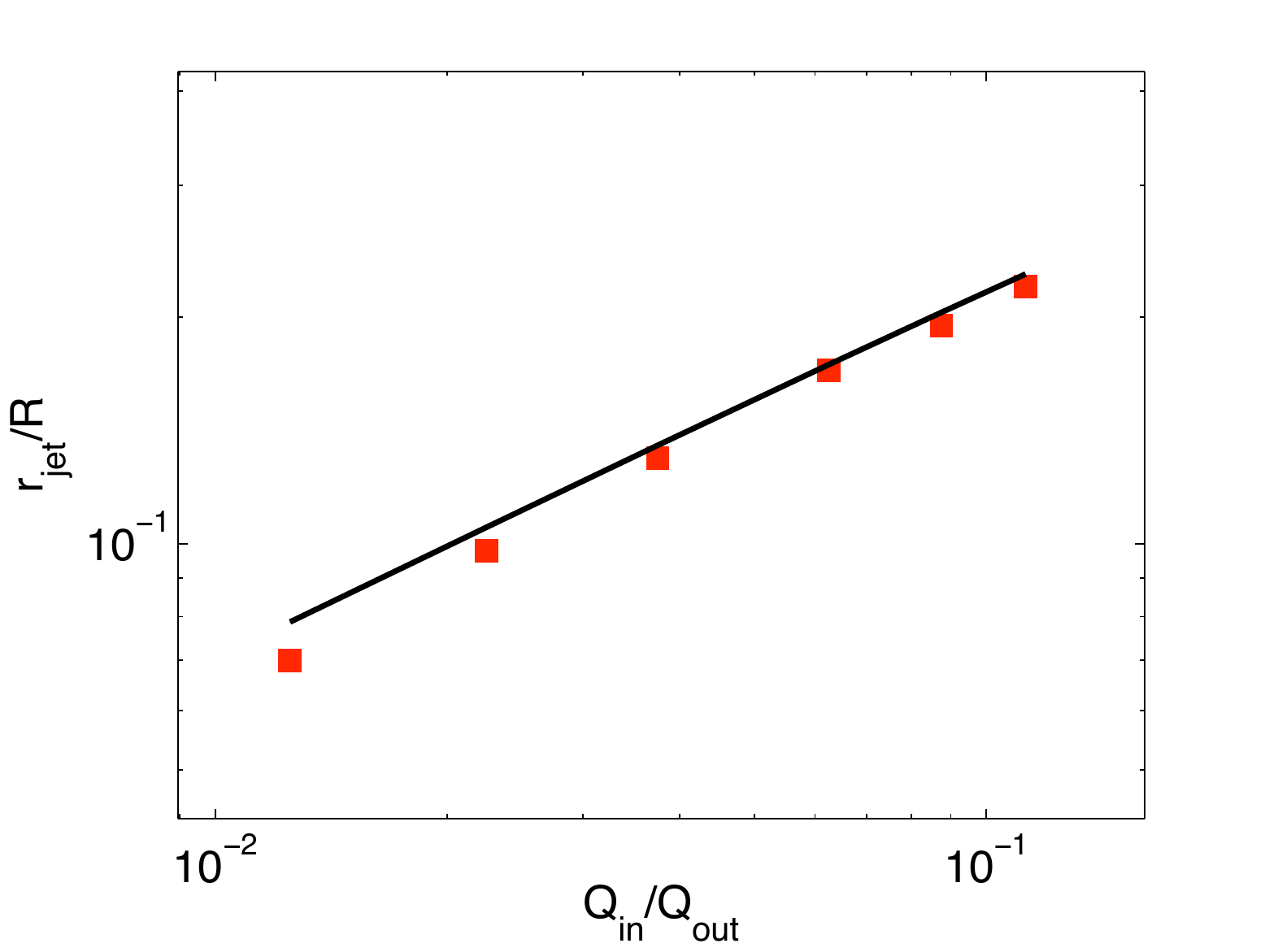}\end{center}
\caption{Diameter of the inner jet rescaled by the diameter of the outer capillary. The black line indicates the scaling law (\ref{eq:scaling_jet}) with a viscosity ratio $\eta_{in}/\eta_{out}=1.25$. The red squares represent the numerical measurements.}
\label{fig:rjet_numerique}
\end{figure}
\end{center}

We have also performed numerical simulations in the regime where an inner jet is produced for a different viscosity ratio, $\eta_{in}/\eta_{out}=1.25$. The resulting measurements collapse well with the scaling law with no adjustable parameters as shown in figure (\ref{fig:rjet_numerique}). 

The scaling law (\ref{eq:scaling_jet}) covers the range of viscosity ratio observed by Utada et al. \cite{utada2007} and is more general as it does not require a low ratio of the inner and outer fluid flow rates. The scaling law is expected to be valid for any flow rates as well as any ratio of viscosity.


\section{Drop formation}

To apply a perturbation on the dispersed flow, we introduce the inlet velocity as follows:
\begin{equation}
w_{in}(t)=w_{m,in}\,\left[1+\epsilon\,\cos(2\pi f t)\right]
\end{equation}

\noindent where $\epsilon$ and $f$ are respectively the amplitude and the frequency of perturbation and $w_{m,in}$ the average inlet velocity. Depending on $\epsilon$ and $f$, different flow morphologies are obtained.

\subsection{Influence of the frequency}

It is known since the work of Rayleigh and Plateau that a jet breaks up into drops due to the interfacial tension effects for a given wavelength of the perturbation \cite{plateau1849,rayleigh1879,eggers2008}. In addition to the spontaneous breakup of the jet at a wavelength chosen by the system, we can facilitate the generation of droplets using an external forcing of the flow at a given frequency. Here, we study the influence of the frequency at a constant amplitude of perturbation, $\epsilon=0.1$ and tune the frequency $f$ in the range $[6\,000\,\text{Hz};20\,000\,\text{Hz}]$. Outside this range of frequencies, no visible effects are observed. Indeed,  it is well-known, from the mechanism responsible for the Rayleigh-Plateau instability, that for perturbation smaller than the radius of the jet, no perturbation can grow and thus the jet does not break up.

\begin{center}
\begin{figure}
\begin{center}\includegraphics[width=10cm]{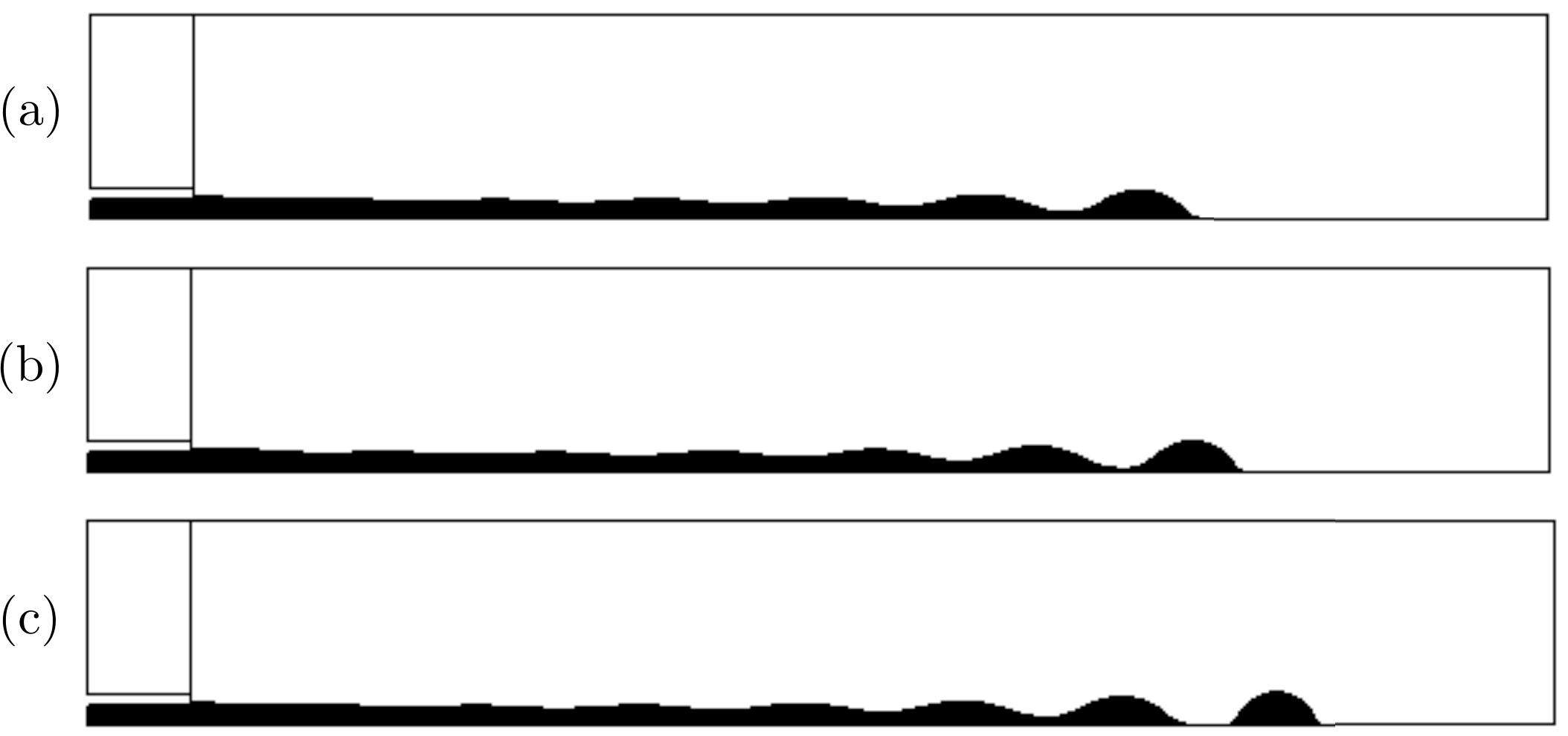}\end{center}
\caption{Morphology of the jet at three successive times for $W\!e_{in}=2.533$, $Ca_{out}=0.3168$, $\eta_{in}/\eta_{out}=1.25$ and $\gamma=20\,\text{mN}\cdot\text{m}^{-3}$. The amplitude and the frequency of perturbation are $\epsilon=0.1$ and $f=16\,000$ Hz, respectively.}
\label{fig:view_16000}
\end{figure}
\end{center}

For a perturbation frequency of $f=16\,000$ Hz, a jetting regime is observed with the production of droplets. We have observed in our numerical simulation that in this case, the generated droplets are monodisperse in size. The morphology of the jets at three consecutive time steps is represented in figure \ref{fig:view_16000}(a), (b), (c). The perturbation on the interface grows along the jet and the radius of the jet becomes periodically thinner with a wavelength $\lambda=2\,\pi/k=2\pi\,v/f$, as shown in figure \ref{fig:view_16000}(a). Then, when the radius becomes very thin, a droplet is generated as exhibited in figure \ref{fig:view_16000}(c).

\begin{center}
\begin{figure}
\begin{center}\includegraphics[width=10cm]{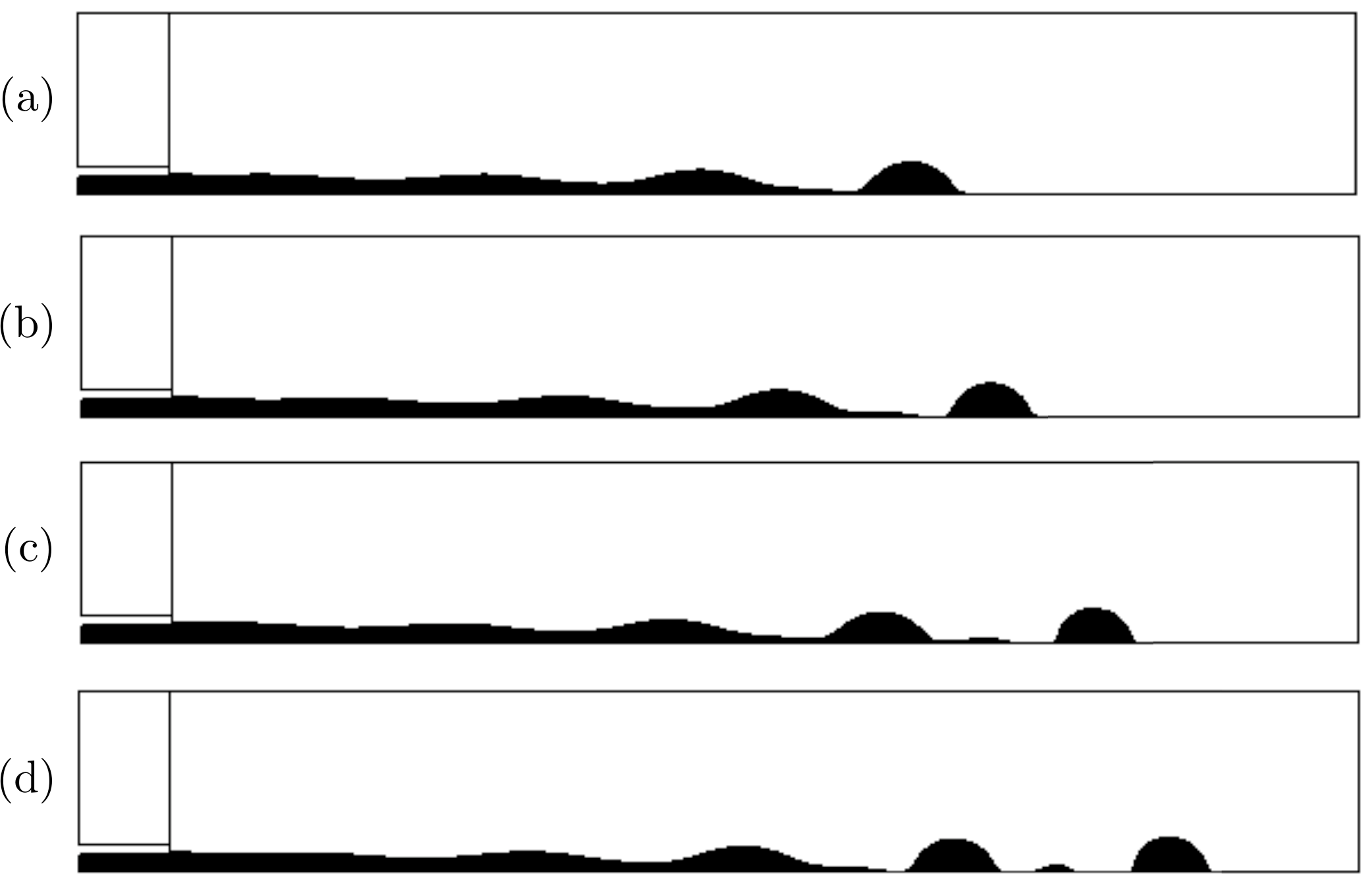}\end{center}
\caption{Morphology of the jet at three successive times for $W\!e_{in}=2.533$, $Ca_{out}=0.3168$, $\eta_{in}/\eta_{out}=1.25$ and $\gamma=20\,\text{mN}\cdot\text{m}^{-3}$. The amplitude and the frequency of perturbation are $\epsilon=0.1$ and $f=10\,000$ Hz, respectively.}
\label{fig:view_10000}
\end{figure}
\end{center}

\begin{center}
\begin{figure}
\begin{center}\includegraphics[width=12cm]{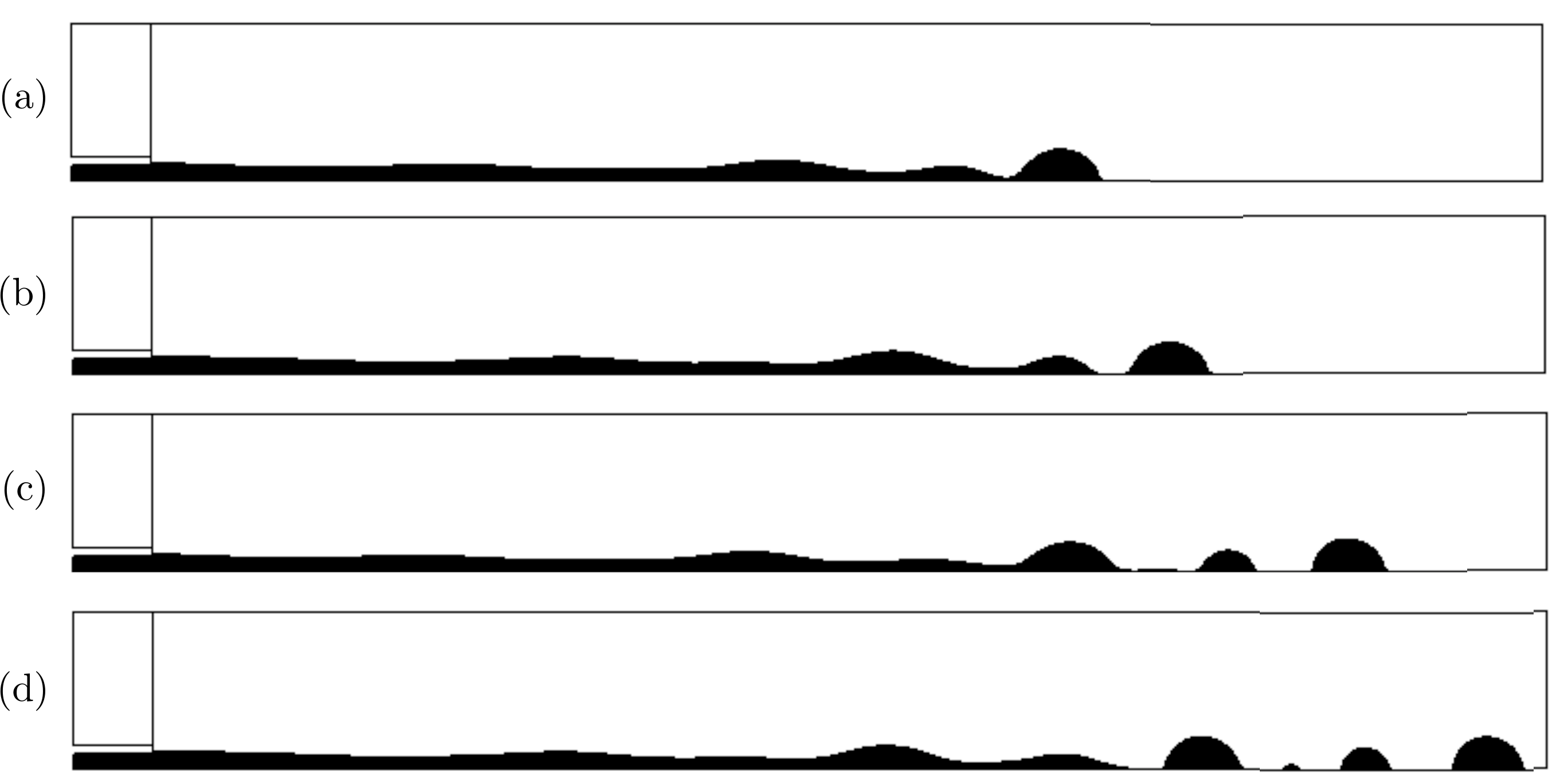}\end{center}
\caption{Morphology of the jet at three successive times for $W\!e_{in}=2.533$, $Ca_{out}=0.3168$, $\eta_{in}/\eta_{out}=1.25$ and $\gamma=20\,\text{mN}\cdot\text{m}^{-3}$. The amplitude and the frequency of perturbation are $\epsilon=0.1$ and $f=6\,000$ Hz, respectively.}
\label{fig:view_6000}
\end{figure}
\end{center}

For a lower frequency, $f=10\,000$ Hz, the dynamics of the breakup remains similar except for the longer pinch-off length due to the larger wavelength $\lambda=2\,\pi\,v/f$. Therefore, the initial breakup occurs for a mother droplet (see figure \ref{fig:view_10000}(a), (b)). An elongated thinner fluid stream later forms a smaller drop called satellite droplet (figure \ref{fig:view_10000}(c), (d)). 

Further decreasing the frequency of perturbation leads to the formation of a mother drop followed by two satellites droplets (figure \ref{fig:view_6000}(a), (b), (c), (d)).

\begin{center}
\begin{figure}
\begin{center}\includegraphics[width=11cm]{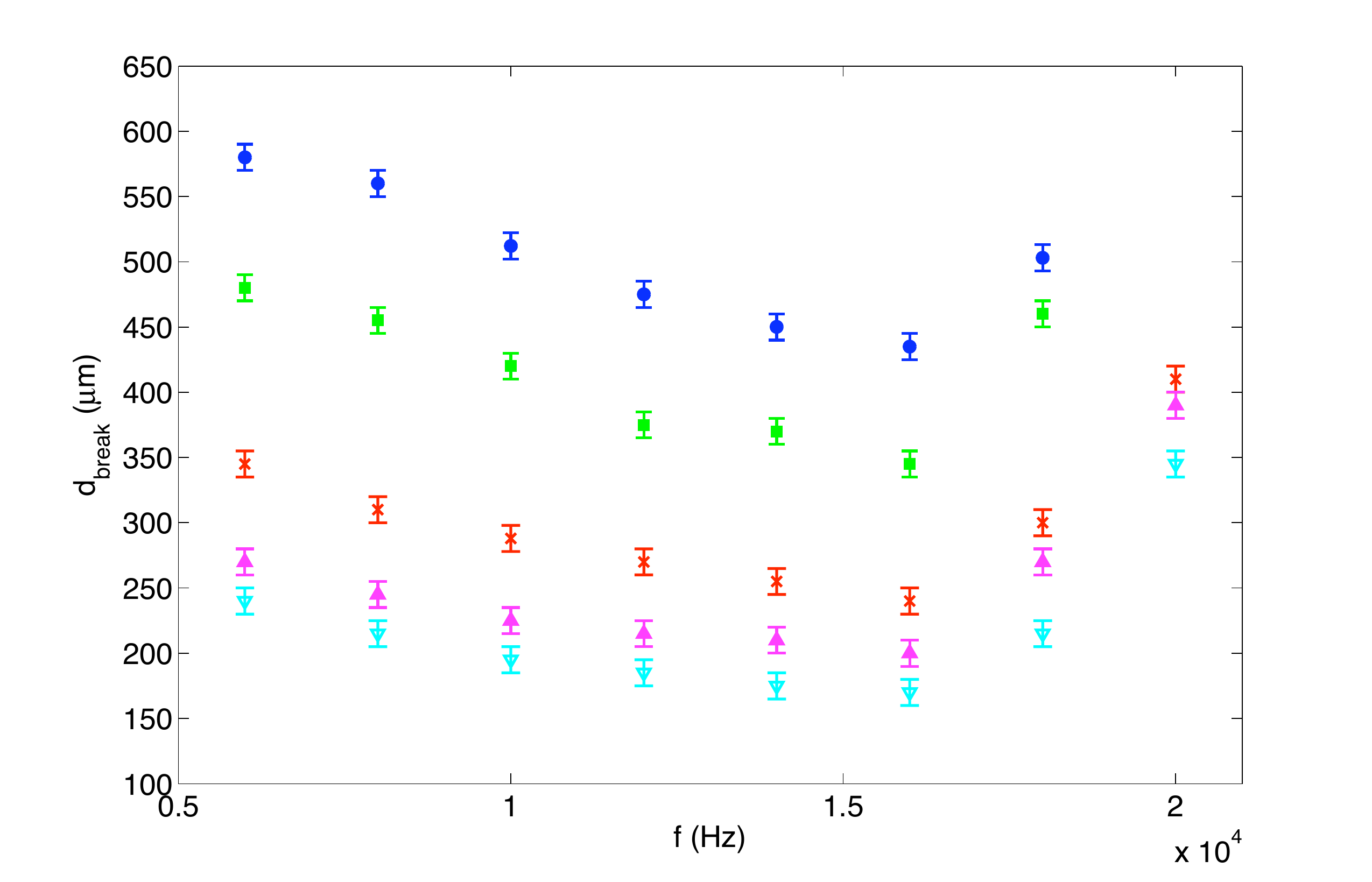}\end{center}
\caption{Length of the jet before breakup (in $\mu m$) as a function of the perturbation frequency for different amplitude: $\epsilon=0.1$ (in blue), $\epsilon=0.2$ (in green), $\epsilon=0.4$ (in red), $\epsilon=0.6$ (in magenta), $\epsilon=0.8$ (in cyan). The inner and outer fluid flow rates are $Q_{in}=1\,800\,\mu\text{L}\cdot\text{h}^{-1}$ and $Q_{out}=80\,000\,\mu\text{L}\cdot\text{h}^{-1}$, respectively.}
\label{fig:dbreak_frequency}
\end{figure}
\end{center}

For given inner and outer fluid flow rates, as well as surface tension and viscosities, the dynamics of the jet is controlled by two key parameters: the frequency and the amplitude of perturbation. We have performed a systematic numerical study to determine the distance at which the jet breakup into droplets as a function of these two parameters $\epsilon$ and $f$. The results are plotted in figure (\ref{fig:dbreak_frequency}). Increasing the amplitude of the initial perturbation leads to a smaller distance for the breakup. An optimal frequency is around $f=16\,000$ Hz in this study but depends on the velocity of the jet.

{{Different theories have been developed to describe the break up of a jet into droplets. In the earlier studies of Rayleigh \cite{rayleigh1879} and Plateau \cite{plateau1849}, an ideal case of an inviscid jet in air is considered; thus the effect of the surrounding fluid is neglected. Tomotika has shown that a surrounding viscous fluid has a non trivial effect on the stability of the jet and consequently, the optimal frequency, where the growth rate of the instability is maximum \cite{tomotika1935}, is modified. In the present study of a coflowing stream in a microcapillary device, both the effect of a surrounding outer fluid and a non-zero velocity of both phases have to be considered. Therefore, we use the theory developed by Guillot et al., \cite{guillot2008a} which considers the dependence of the breakup on the wavenumber of the perturbation, $k$, for a co-flow jet}}. They found that for a perturbation of the form,
\begin{equation}
r=r_0(1+\epsilon\,\text{e}^{\text{i}\omega t})
\end{equation}
a dispersion relation between $\omega$ and $k$ can be derived
\begin{equation}
\omega=-\text{i}\,K_a\,\frac{E(x)}{x^2}\,k+\frac{F(x)}{x^5}(k^2-k^4)
\end{equation}
where 
\begin{eqnarray}
E(x)=\frac{d(x)\,b(x)-a(x)\,e(x)}{a(x)+d(x)}\nonumber \\
F(x)=\frac{c(x)\,d(x)-f(x)\,e(x)}{a(x)+d(x)}\nonumber
\end{eqnarray}
\noindent with
\begin{eqnarray}
a(x) & =  & -(1-x^2)^2 \quad\qquad\qquad\qquad b(x)=4\,x\,(1-x^2) \nonumber \\
c(x)& = & 2\,x^4-2\,x^2-4\,x^4\,\text{ln}(x) \qquad d(x)=2\,x^2\,(x^2-1)-\frac{\eta_{out}}{\eta_{in}}\,x^4 \nonumber \\
e(x) & = & 8\,x^3-4\,\frac{\eta_{out}}{\eta_{in}}\,x^3-4\,x \quad\qquad f(x)=4\,x^4\,\text{ln}(x)-\frac{\eta_{out}}{\eta_{in}}\,x^4 \nonumber
\end{eqnarray}

The capillary number based on the capillary radius and the ratio of radii are defined, respectively, by
\begin{eqnarray}
K_a=\frac{-\delta_z P\,R^2}{\gamma} \\
x=\frac{r_j}{R}
\end{eqnarray}

These two dimensionless numbers are related to the inner and outer fluid flow rates through the relations
\begin{eqnarray}
Q_{out}=\frac{\pi\,\gamma\,R^2}{8\,\eta_{out}}\left[1-x^2\right]^2\,K_a \\
Q_{in}=Q_{out}\left[\frac{\eta_{out}}{\eta_{in}}\,\frac{x^4}{(1-x^2)^2}+\frac{2\,x^2}{1-x^2}\right]
\end{eqnarray}

Therefore, for an imposed wavenumber $k$ related to $f$ through the relation
\begin{equation}
k=\frac{2\,\pi\,v}{f}
\end{equation}

The growth rate of the perturbation is given by the real part of $\omega=\omega_r$:
\begin{equation}
\omega_r=\frac{F(x)}{x^5}(k^2-k^4)
\end{equation}

\begin{center}
\begin{figure}
\begin{center}\includegraphics[width=8cm]{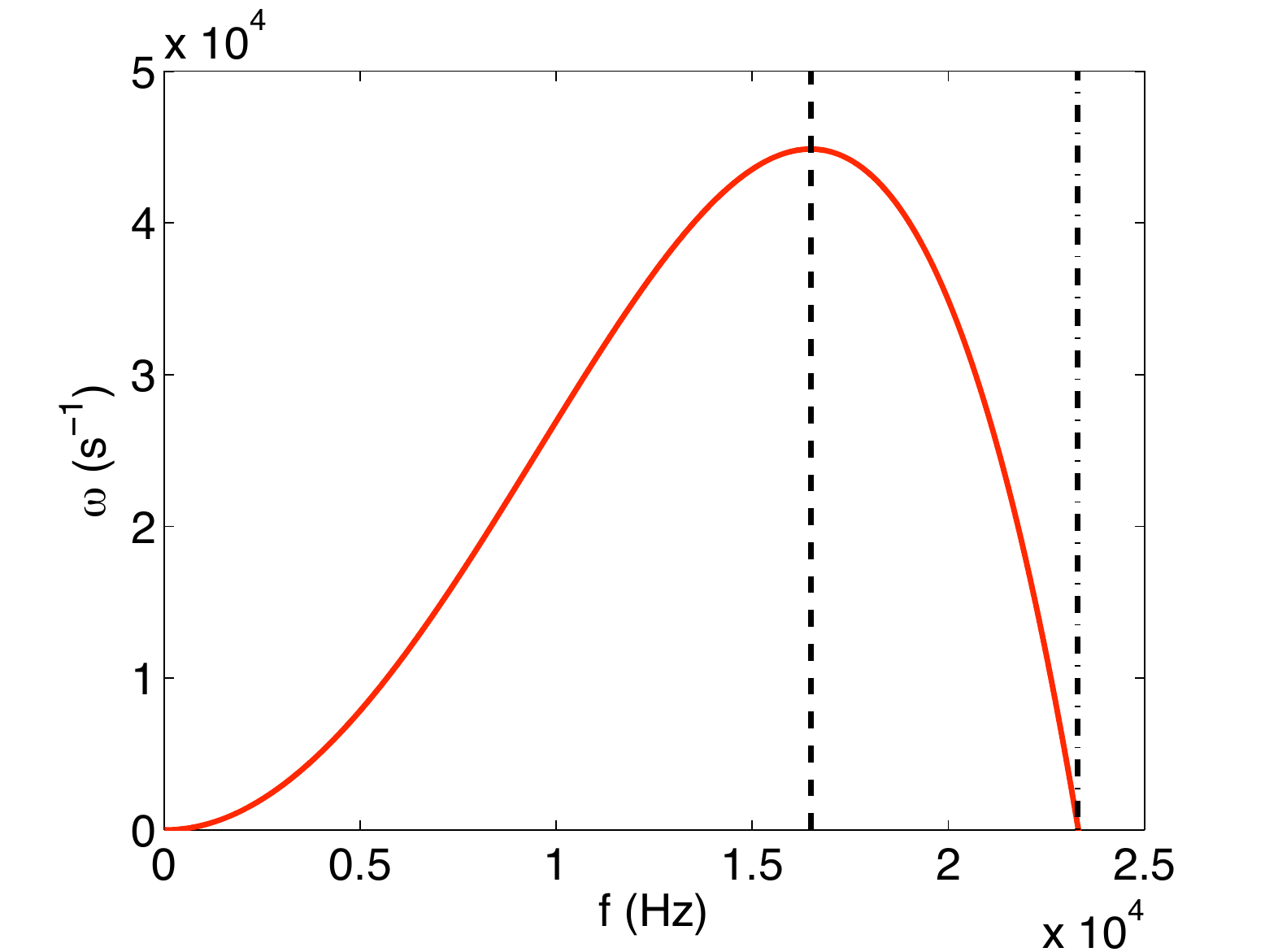}\end{center}
\caption{Growth rate of the Rayleigh-Plateau instability $\omega_r$ (in $s^{-1}$) for an imposed perturbation frequency $f$ (in Hz). The dashed-dotted line located around $f=23\,000$ Hz indicates the critical frequency beyond which no perturbation can grow. The black dotted line around $f=16\,000$ Hz indicates the frequency where the growth rate is maximum.}
\label{fig:growth_rate}
\end{figure}
\end{center}

The growth rate $\omega_r$ as a function of the perturbation frequency is plotted in figure (\ref{fig:growth_rate}). It has been shown using energetic consideration that modes whose wavelength $2\pi/k$ are smaller than the perimeter of the jet have a negative difference in energy between the perturbed and the unperturbed jet. Therefore, no perturbation with wavenumber $k>1/r_0$ can grow \cite{eggers2008}. This is in agreement with our numerical simulation where no growth of perturbation for $f > 23\,000$ Hz can be observed; this corresponds to $k\,r_{jet}>1$ (see also figure (\ref{fig:growth_rate})). The optimal frequency of perturbation to break up the jet into drops at the present flow rate ($Q_{in}=1\,800\,\mu\text{L}\cdot\text{h}^{-1}$ and $Q_{out}=80\,000\,\mu\text{L}\cdot\text{h}^{-1}$) is $f\simeq16\,000$ Hz, in agreement with the results from numerical simulation (see figure \ref{fig:dbreak_frequency}). This theoretical description is thus well-adapted to describe the destabilization of a co-flow jet and therefore provides a criterion to choose experimentally the optimal frequency to break up a co-flow jet into droplets using induced perturbations.

\subsection{Scaling of the droplets size}

For given inner and outer fluid flow rates, the generation of monodisperse droplets is possible in a range of perturbation frequency. For $Q_{in}=1\,800\,\mu\text{L}\cdot\text{h}^{-1}$ and $Q_{out}=80\,000\,\mu\text{L}\cdot\text{h}^{-1}$, this is achievable for $f \in[6\,000;18\,000]$ Hz as exhibited in figure (\ref{fig:dbreak_frequency}). However, the size of the resulting drops together with the number of satellite droplets varies with the applied frequency (see figures \ref{fig:view_16000}, \ref{fig:view_10000}, \ref{fig:view_6000}). The inability to predict the number and volumes of the satellite droplets limit the ability to estimate the exact radius of the main emulsion droplets. Therefore, we consider the total volume of fluid contained in all drops as
\begin{equation}
V_d=\frac{4}{3}\,\pi\left[{r_{drop}}^3+\sum_i{r_{sat,i}}^3\right]
\end{equation}
\noindent where the first term accounts for the mother drops and the second term for the satellite droplets. In some cases, more than one satellite droplets are present, the symbol $i$ denotes the particular satellite droplet and its radius $r_{sat,i}$. The volume of the jet, which is a sum of the volume of all satellite droplets, writes
\begin{equation}
V_{jet}=\pi\,{r_{jet}}^2\,\frac{w_{jet}}{f}=V_d
\end{equation}
\noindent where $\pi\,{r_{jet}}^2$ is the transverse area of the jet and $w_{jet}/f$ is the periodic distance over one perturbation. Moreover, the inner flow rate can be expressed as:
\begin{equation}
Q_{in}=\pi\,{r_{jet}}^2\,w_{jet}
\end{equation}
\noindent leading to a simple relation for the volume of one main drop and its satellite droplets:
\begin{equation}\label{scale_ouhou}
V_d=\frac{4}{3}\,\pi\left[{r_{drop}}^3+{r_{sat}}^3\right]=\frac{Q_{in}}{f}
\end{equation}

\begin{center}
\begin{figure}
\begin{center}\includegraphics[width=9cm]{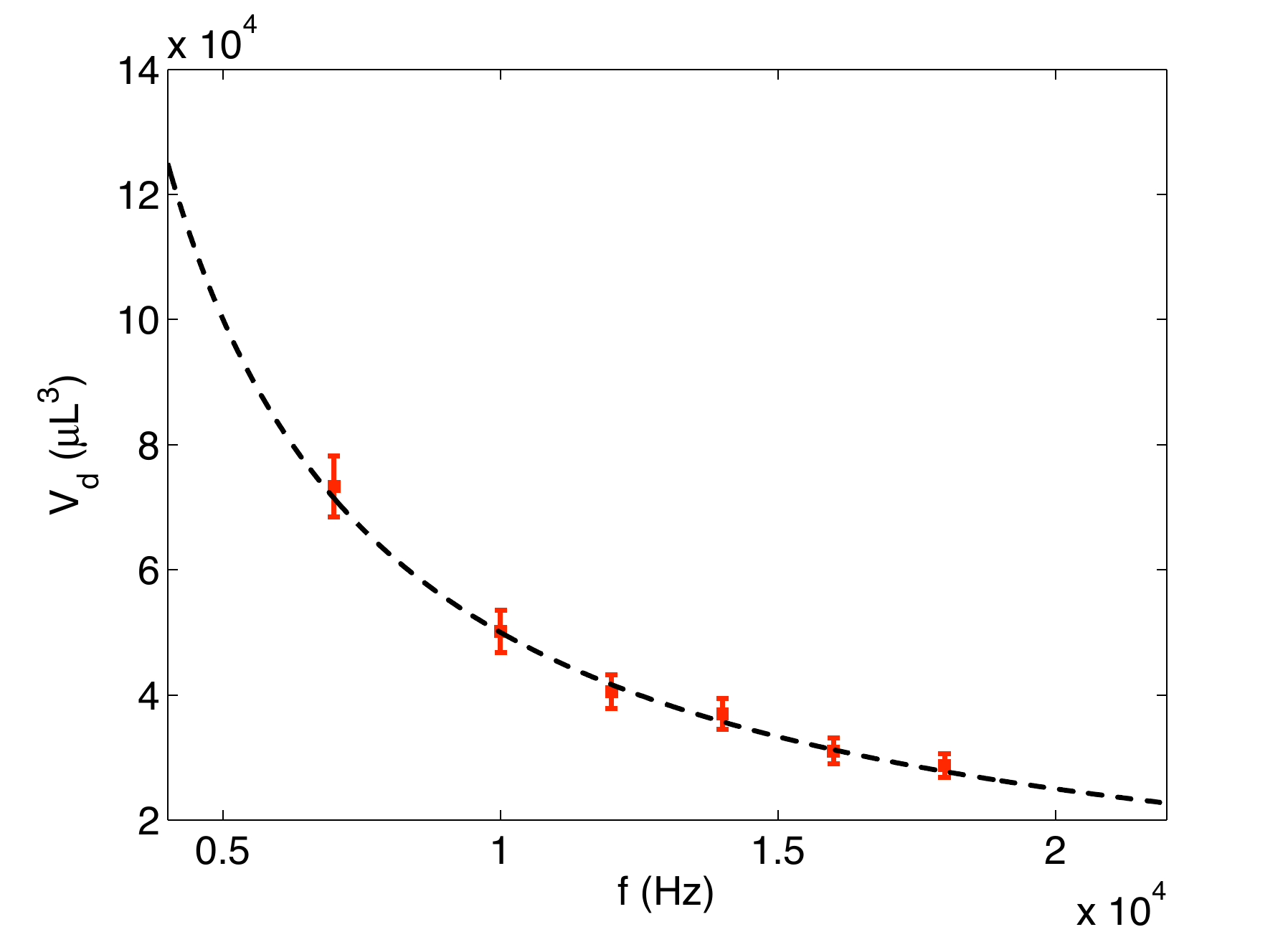}\end{center}
\caption{Volume encapsulated in the mother and satellites drops $V_d$ (in $\mu\text{L}^{3}$) as a function of the perturbation frequency $f$ (in Hz). The red squares are measurements using numerical simulation together with the uncertainties and the dashed black line is the scaling law (\ref{scale_ouhou}) without any adjustment parameters. The inner and outer fluid flow rates are $Q_{in}=1\,800\,\mu\text{L}\cdot\text{h}^{-1}$ and $Q_{out}=80\,000\,\mu\text{L}\cdot\text{h}^{-1}$, respectively.}
\label{fig:scaling_droplet}
\end{figure}
\end{center}

This relation shows that the encapsulated volume of fluid does not depend on the surface tension $\gamma$ or on the viscosity. We measure the radius of the different drops obtained in the numerical simulation for inner and outer flow rates of $Q_{in}=1\,800\,\mu\text{L}\cdot\text{h}^{-1}$ and $Q_{out}=80\,000\,\mu\text{L}\cdot\text{h}^{-1}$ respectively, with perturbation with an amplitude of $\epsilon=0.1$ and frequencies in the range $f\in[6000;18000]$. The resulting measurements show a very good agreement with the scaling law derived without any adjustable parameters, as shown in figure \ref{fig:scaling_droplet}.

\subsection{Influence of the amplitude of perturbation}

For a given perturbation frequency, the breakup length show a clear dependence on the initial perturbation amplitude (see figure \ref{fig:dbreak_frequency}). Considering an initial perturbation of the jet radius such that:
\begin{equation}
r_{jet}(t)=r_{jet}\left[1-\epsilon_0\,\text{e}^{\omega\,t}\right]
\end{equation}
the jet can be considered as broken up into droplets at $t_0$ for $r_{jet}(t_0) \sim 0$. This leads to a condition on $\epsilon_0$
\begin{equation}
\epsilon_0\,\text{e}^{\omega\,t_0}\sim 1
\end{equation}
The distance $d_{break}$ for the jet to break up is related to the time $t_0$ and to the velocity of the jet $w_{jet}$ through the relation $t_0=d_{break}/w_{jet}$. It follows a relation between the initial disturbance amplitude $\epsilon_0$ and the distance $d_{break}$:
\begin{equation}\label{eq:scalin88}
d_{break} \sim -\frac{v_{jet}}{\omega_r}\,\text{ln}\,\epsilon_0
\end{equation}

\begin{center}
\begin{figure}
\begin{center}\includegraphics[width=11cm]{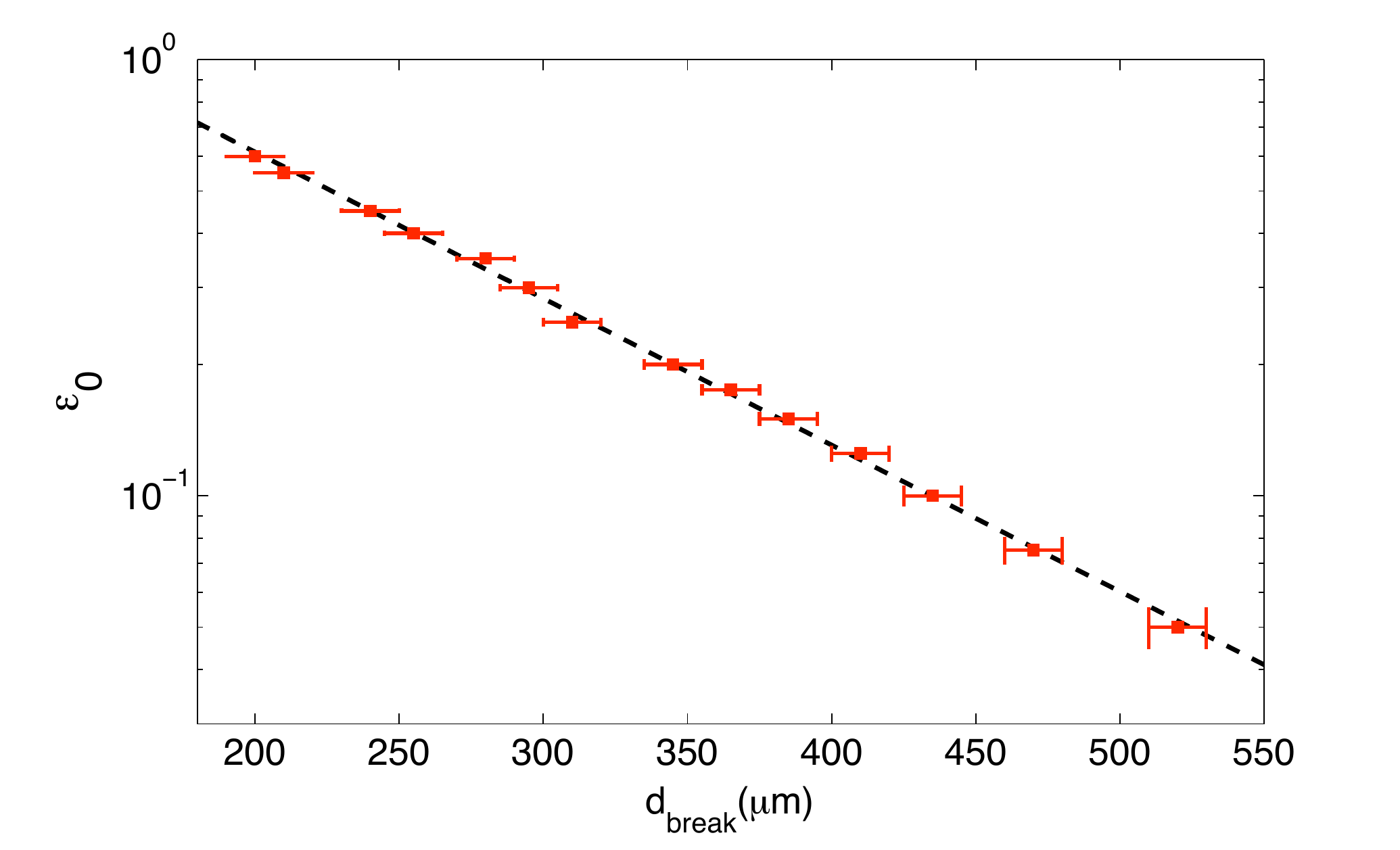}\end{center}
\caption{Distance for the jet to break up into drops, $d_{break}$ (in $\mu m$), as a function of the initial amplitude of perturbation $\epsilon_0$ (red squares) together with the scaling law (\ref{eq:scalin88}) (dotted line).}
\label{fig:epsilon_length_scaling}
\end{figure}
\end{center}

For given inner and outer fluid flow rates together with fixed interfacial tension and viscosities, $v_{jet}$ and $\omega_r$ remain constant and thus a simple scaling law is obtained. We measure, in the numerical simulation, the distance $d_{break}$ for a perturbation frequency $f=16\,000$ Hz, $Q_{in}=1\,800\,\mu\text{L}\cdot\text{h}^{-1}$ and $Q_{out}=80\,000\,\mu\text{L}\cdot\text{h}^{-1}$. The results from the numerical simulation agree well with the scaling law, as plotted in figure (\ref{fig:epsilon_length_scaling}).

\section{Conclusion}

In this paper, we study the dynamics of a co-flow jet in the regime where usually the jet does not break up into monodisperse droplets. Using an axisymmetric numerical simulation, we show that the introduction of an initial perturbation to the velocity of the dispersed phase can lead to a faster breakup of the jet with a good control over the size of the droplets. A scaling law is derived for the radius of the unperturbed jet in a co-flow stream; the prediction from the scaling law agrees well with previous experimental results in the literature and numerical measurements. The two controlable parameters, the frequency $f$ and the amplitude $\epsilon$ of the perturbation, are shown to have a strong effect on the flow. A scaling law for the resulting volume of dispersed phase contained in the droplets is provided and the distance to break the jet is characterized with respect to the frequency. 

The new results lead to a better understanding of the fluid dynamics of a perturbed jet; this is particulary helpful in the new field of water-in-water emulsion where a perturbation in the pressure or velocity field must be introduced to form monodisperse drops. We show that the optimal frequency can be estimated using theoretical consideration and can be directly applied to the numerical simulations or in experiments.

By providing guidelines on how to break up jets into monodisperse droplets with controlled sizes, our work suggests new strategies to extend the applicability of droplet microfluidic techniques to a wider variety of fluidic systems.

\bibliography{biblio}

\end{document}